\documentclass [aps,prb,preprint,floatfix,showpacs,superscriptaddress]{revtex4}

\usepackage {amsmath,amssymb,epsfig,color,tabularx}

\begin{document}

\title
{Orbital densities functional}

\author{V.~I.~Anisimov}
\affiliation{Institute of Metal Physics, Russian Academy of
Sciences-Ural Division, 620041 Yekaterinburg GSP-170, Russia}
\author{A.~V.~Kozhevnikov}
\affiliation{Institute of Metal Physics, Russian Academy of
Sciences-Ural Division, 620041 Yekaterinburg GSP-170, Russia}
\author{M.~A.~Korotin}
\affiliation{Institute of Metal Physics, Russian Academy of
Sciences-Ural Division, 620041 Yekaterinburg GSP-170, Russia}
\author{A.~V.~Lukoyanov}
\affiliation{Institute of Metal Physics, Russian Academy of
Sciences-Ural Division, 620041 Yekaterinburg GSP-170, Russia}
\affiliation{Ural State Technical University-UPI,
620002 Yekaterinburg, Russia}
\author{D.~A.~Khafizullin}
\affiliation{Institute of Metal Physics, Russian Academy of
Sciences-Ural Division, 620041 Yekaterinburg GSP-170, Russia}
\affiliation{Ural State Technical University-UPI,
620002 Yekaterinburg, Russia}

\date{\today}
\pacs {78.70.Dm, 71.25.Tn}

\begin{abstract}
Local density approximation (LDA) to the density functional theory (DFT) 
has continuous derivative of total energy as a number of electrons 
function and continuous exchange-correlation potential, 
while in \emph{exact} DFT both should be discontinuous as number 
of electrons goes through an integer value. We propose orbital 
densities functional (ODF) (with orbitals defined as Wannier functions) 
that by construction obeys this discontinuity condition. By its variation 
one-electron equations are obtained with potential in the form of projection 
operator. The operator increases a separation between occupied and empty 
bands thus curing LDA deficiency of energy gap value systematic 
underestimation. Orbital densities functional minimization gives ground 
state orbital and total electron densities. The ODF expression for the energy 
of orbital densities fluctuations around the ground state values defines 
ODF fluctuation Hamiltonian that allows to treat correlation effects. 
Dynamical mean-field theory (DMFT) was used to solve this Hamiltonian 
with quantum Monte Carlo (QMC) method for effective impurity problem. 
We have applied ODF method to the problem of metal-insulator transition 
in lanthanum trihydride LaH$_{3-x}$. In LDA calculations ground state 
of this material is metallic for all values of hydrogen nonstoichiometry 
$x$ while experimentally the system is insulating for $x<0.3$. ODF method 
gave paramagnetic insulator solution for LaH$_3$ and LaH$_{2.75}$ 
but metallic state for LaH$_{2.5}$.

\end{abstract}
\maketitle
\section{Introduction}
Numerical electronic structure calculations are now well
established branch of solid state physics. While for the finite systems,
such as atoms and molecules, more sophisticated and rigorous calculation
methods exist, for extended systems studied in condensed matter physics the
only widely used practical tool is up to now density functional theory (DFT)
in the local density approximation (LDA).~\cite{HK,Kohn-Sham} It has so great 
predictive power, that not only charge and spin density, one-electron and total 
energies obtained in LDA are in general in a very good agreement with experimental 
data, but it was possible to develop \emph{ab initio} molecular dynamic methods, 
based on LDA. Such methods achieved the level of numerical experiment, because
even so complicated effects, such as reconstruction of the crystal surface can
be correctly described by them.~\cite{LDA_Gen}

However, there are materials where LDA results do not agree well with experimental data. 
For band insulators and semiconductors
LDA gives systematically underestimated values of the energy
gap.~\cite{jones} For Mott insulators, for example transition-metal
oxides, LDA results could be qualitatively wrong,
giving metallic state while experimentally those systems are wide
gap insulators.~\cite{cuprates}

There were many attempts to cure this deficiency of LDA. Among the
most widely used one can mention GW,~\cite{GW} SIC,~\cite{SIC} and
LDA+$U$\cite{ani91,ani97} methods. While those approaches have their
advantages there is still no universally accepted calculation
scheme which would be as simple and practical as standard LDA and
a search for better methods continues in scientific community.

The basic problems of LDA can be traced to the fact that the exchange-correlation 
energy functional is defined in a \emph{local} approximation. As a result its 
variational derivative, exchange-correlation potential (that is a \emph{function} 
of the density value in particular point \textbf{r} instead 
of being a general \emph{functional} of density), is continuous as a number 
of electrons function. More general approximations using in addition to the electron density 
its gradient like GGA \cite{GGA} method have continuous potential as well. 
However, Perdew {\it et al. }\cite{perdew} had investigated the properties 
of the \emph{exact} density functional (EDF) and shown that its potential 
must jump discontinuously when number of electrons $N$ goes through 
an integer value. The proper function of total energy $E$ versus number 
of electrons $N$ should have a curve as series
of straight-line segments with derivative discontinuities at
integral values of $N$ while in LDA this curve has continuous derivatives. 
Any attempts to improve LDA as an approximation to exact density functional 
theory should be done in such a way that a new functional would obey this discontinuity requirement. 

In the present work we define a functional that by construction has the potential 
discontinuities required for exact density functional. For that we introduce 
a concept of \emph{orbital densities} corresponding to one-electron orbitals. 
The orbital densities functional (ODF) depends on a set of \emph{orbital densities} 
instead of the total electron density only. When the number of electrons goes through 
an integer value the variational derivative of the ODF functional jumps discontinuously 
and the corresponding function of total energy $E_{ODF}$ versus number of electrons $N$ 
has a curve as series of straight-line segments. 

Varying this functional one-electron equations were obtained with potential in the form 
of projection operator. For occupied states the effect of this operator on the energy 
is negative and positive for empty states. As a the result it increases separation 
between valence and conduction bands comparing with LDA values thus curing LDA deficiency 
of energy gap value systematic underestimation. 

The minimization of ODF functional gives a set of orbital densities and hence the total
electron density corresponding to ground state of the system. The same functional can be 
used to calculate energies of orbital densities fluctuations around the ground state values. 
From this we have derived \emph{fluctuation Hamiltonian} defined via orbital density 
fluctuation operators. This Hamiltonian allows to treat correlation effects and hence 
obtain better description of ground state properties and spectral function for the excitations. 
In the present work we have used for ODF Hamiltonian dynamical mean-filed theory 
(DMFT)~\cite{vollha93,pruschke,georges96} with quantum Monte Carlo (QMC) method 
to solve the effective impurity problem.

The optimal choice for one-electron orbitals needed to define orbital densities can be 
determined by condition of fluctuation energy minimum. We have shown that the less extended 
in space these orbitals are, the lower the energy of orbital densities fluctuations 
around ground state is. Therefore in our ODF method we have used maximally localized 
Wannier functions for orbital densities definition.~\cite{vanderbildt}

Recently we have developed ``generalized transition state'' (GTS) method \cite{GTS} 
to improve agreement of calculated and experimental spectral properties comparing with LDA. 
We have found that ODF projection operator is identical to the one in equations of GTS method. 
The basis of GTS method was an idea that one-electron energies corresponding to Wannier 
functions should have a meaning of the removal (addition) energies for electrons from 
(to) the corresponding states. That was realized by using ``transition state'' 
scheme\cite{slater} generalized to Wannier functions. It is remarkable that so different 
approaches as ``transition state'' correction to excitation energies and requirement 
of exchange-correlation potential discontinuity can lead to the same equations.

We have applied ODF method to the problem of metal-insulator transition in LaH$_{3-x}$. 
LDA has severe difficulties for this material because it gives metallic solution for all 
values of hydrogen deficiency $x$ while experimentally the system is insulating for $x<0.3$. 
We have found that ODF projection operator potential is enough to open a gap for 
stoichiometric LaH$_3$ but in order to reproduce paramagnetic insulator for LaH$_{2.75}$ 
correlation effects should be taken into account via DMFT-QMC solution of ODF 
fluctuation Hamiltonian. 

The paper is structured as follows. In Sec.~\ref{ODF} the derivations of ODF functional 
and fluctuation Hamiltonian are  presented. In Sec.~\ref{W_def} and~\ref{WF_GF} we
describe the construction of Wannier functions. Sec.~\ref{WF_DMFT} gives a short account 
of DMFT method and Sec.~\ref{self-cons} describes the self-consistency issues. 
In Sec.~\ref{LaH3} results of ODF calculations for LaH$_{3-x}$ are presented. 
Sec.~\ref{Conclusion} concludes the paper.

\section{Orbital densities functional and fluctuation Hamiltonian}
\label{ODF}
In the local density approximation (LDA)\cite{jones} 
to the density functional theory \cite{Kohn-Sham}(see Appendix~\ref{constrain})
exchange-correlation energy is calculated via:
\begin{eqnarray}
\label{ODF2} E^{LDA}_{xc}[\rho]=\int d\mathbf{r}
\varepsilon_{xc}(\rho(\mathbf{r}))\rho(\mathbf{r}),
\end{eqnarray}
where $\varepsilon_{xc}(\rho(\mathbf{r}))$ is exchange-correlation
energy density for homogeneous electron gas with density equal to
$\rho(\bf{r})$.

Equation (\ref{ODF2}) defines an exchange-correlation potential
$V_{xc}(\rho(\mathbf{r}))=\delta(E_{xc}[\rho])/\delta\rho(\mathbf{r})$ which is a
continuous function of the number of electrons $N$. However, Perdew
{\it et al.}~\cite{perdew} proved that for the {\it{exact}} density
functional exchange-correlation potential must jump discontinuously
when number of electrons $N$ goes through an integer value.

The Hohenberg-Kohn theorem\cite{HK} was extended in Ref.~\onlinecite{perdew} 
to fractional electron number. It was shown that for the electron density
$\rho(\mathbf{r})$ which integrates to $N=M+\omega$ where M is an
integer and $0\leq\omega\leq1$ the exact density functional
$E_{EDF}$ is equal to:
\begin{eqnarray}
\label{LDA23} E_{EDF}(M+\omega)=(1-\omega)E_{EDF}(M)+\omega
E_{EDF}(M+1).
\end{eqnarray}
This means that in general the curve of $E_{EDF}$ versus $N$ is a series
of straight-line segments with derivative discontinuities at
integer values of $N$.

It was proven\cite{perdew} that the chemical potential 
$\mu={\partial E_{EDF}}/{\partial N}$ is discontinues when 
the number of electrons goes through integer value:
\begin{eqnarray}\label{LDA25}
\mu = \left\{\begin{array}{ll}
-I &   \, (M-1 < N < M) \\
-A &    \, (M < N < M+1)
\end{array} \right.
\end{eqnarray}
where $I=E_{EDF}(M-1)-E_{EDF}(M)$ and $A=E_{EDF}(M)-E_{EDF}(M+1)$
are removal and addition energies respectively. The functional
derivative $\delta E_{EDF}/\delta \rho(\mathbf{r})$ is
also discontinuous: two limits for $N$ approaching $M$ from above
and below will differ by constant $I-A$.

Electron density [Eq.~(\ref{LDA43})] can be expressed as a sum of
the ``orbital densities'' $\rho_i(\mathbf{r})$ defined in the
following way ($\psi_i(\bf{r})$ is $i$-th orbital wave function):
\begin{eqnarray}
\label{ODF1} \rho(\mathbf{r}) = \sum_i \rho_i(\bf{r}), \\ \nonumber
\rho_i(\mathbf{r})\equiv n_i|\psi_i(\bf{r})|^2.
\end{eqnarray}
The orbital densities can be varied from zero to the maximum
values $\rho^{max}_i(\mathbf{r})=|\psi_i(\bf{r})|^2$.

The condition of linear dependence of the {\it{exact}} density
functional on the fractional number of electrons
[Eq.~(\ref{LDA23})] can be expressed via orbital density
$\rho_j(\mathbf{r})$ corresponding to the partially occupied 
orbital $j$. This orbital is the lowest unoccupied one for number 
of electrons $N=M$ and the highest occupied for $N=M+1$. 
The variation of the total electron density
$\rho(\mathbf{r})$ will be defined only by the variation of $\rho_j(\mathbf{r})$
for number of electrons changing from $M$ to $M+1$
\begin{eqnarray}
\label{ODF3}
E_{EDF}[\rho]=E_{EDF}[\rho_0+\rho_j]=E_{EDF}\Big|_{\rho_j(\mathbf{r})=0}+ \int
d\mathbf{r} \rho_j(\mathbf{r})\Lambda_j(\mathbf{r}),
\end{eqnarray}
where $\rho_0(\mathbf{r})$ is the density for number of electrons $N=M$.
Here $\Lambda_j(\mathbf{r})$ satisfies to the equation:
\begin{eqnarray}
\label{ODF4} \int d\mathbf{r}
\rho^{max}_j(\mathbf{r})\Lambda_j(\mathbf{r})=
E_{EDF}\Big|_{\rho_j(\mathbf{r})=\rho^{max}_j(\mathbf{r})}-E_{EDF}\Big|_{\rho_j(\mathbf{r})=0}.
\end{eqnarray}

Orbital density $\rho_j(\mathbf{r})$ enters Eq.~(\ref{ODF3}) in
a linear form. However LDA functional [Eqs.~(\ref{LDA44}) and (\ref{ODF2})] does not
show such a linear dependence on density variations. Equation
analogous to Eq.~(\ref{ODF3}) for LDA has a general form (keeping
only first and second variational derivatives in expansion series):
\begin{eqnarray}
\label{ODF5} E_{LDA}[\rho]=E_{LDA}[\rho_0+\rho_j]  \approx
E_{LDA}\Big|_{\rho_j(\mathbf{r})=0}+ \int d\mathbf{r}
\rho_j(\mathbf{r}){\frac{\delta
E_{LDA}}{\delta\rho(\mathbf{r})}}\Big|_{\rho_j(\mathbf{r})=0}
\\ \nonumber  + \frac{1}{2}\int d\mathbf{r} \rho_j(\mathbf{r})\int
d\mathbf{r'}\rho_j(\mathbf{r'})\frac{\delta^2
E_{LDA}}{\delta\rho(\mathbf{r})\delta\rho(\mathbf{r'})}\Big|_{\rho_j(\mathbf{r})=0}.
\end{eqnarray}

Using Eqs.~(\ref{LDA60}) and (\ref{LDA61}) this expansion can be rewritten via constrain potential $\Phi(\mathbf{r})$ and effective interaction strength function $W(\mathbf{r},\mathbf{r'})$:
\begin{eqnarray}
\label{ODF5s}E_{LDA}[\rho]=E_{LDA}[\rho_0+\rho_j]  \approx
E_{LDA}\Big|_{\rho_j(\mathbf{r})=0}+ \int d\mathbf{r}
\rho_j(\mathbf{r})\Phi(\mathbf{r})\Big|_{\rho_j(\mathbf{r})=0}
\\ \nonumber  + \frac{1}{2}\int d\mathbf{r} \rho_j(\mathbf{r})\int
d\mathbf{r'}\rho_j(\mathbf{r'})W(\mathbf{r},\mathbf{r'})\Big|_{\rho_j(\mathbf{r})=0}.
\end{eqnarray}

If one will add to Eq.~(\ref{ODF5}) a correction term:
\begin{eqnarray}
\label{ODF6} E_{corr}[\rho_j]  \equiv -\frac{1}{2}\int
d\mathbf{r} \rho_j(\mathbf{r})\int
d\mathbf{r'}(\rho_j(\mathbf{r'})-\rho^{max}_j(\mathbf{r'}))W(\mathbf{r},\mathbf{r'})\Big|_{\rho_j(\mathbf{r})=0}
\end{eqnarray}
then the linear dependence on the density variations (as it is required for the
{\it{exact}} density functional [Eq.~(\ref{ODF3})]) will be restored:
\begin{eqnarray}
\label{ODF7} E_{LDA}[\rho_0+\rho_j]+
E_{corr}[\rho_j] \approx
E_{LDA}\Big|_{\rho_j(\mathbf{r})=0}
\\+ \nonumber \int d\mathbf{r} \rho_j(\mathbf{r})\Big(\Phi(\mathbf{r})\Big|_{\rho_j(\mathbf{r})=0}+
\frac{1}{2}\int
d\mathbf{r'}\rho^{max}_j(\mathbf{r'})W(\mathbf{r},\mathbf{r'})\Big|_{\rho_j(\mathbf{r})=0}\Big).
\end{eqnarray}
Equation (\ref{ODF7}) is equivalent to Eq.~(\ref{ODF3}) with the function
$\Lambda_j(\mathbf{r})$ equal to:
\begin{eqnarray}
\label{ODF8s} \Lambda_j(\mathbf{r})=\Phi(\mathbf{r})\Big|_{\rho_j(\mathbf{r})=0}+
\frac{1}{2}\int
d\mathbf{r'}\rho^{max}_j(\mathbf{r'})W(\mathbf{r},\mathbf{r'})\Big|_{\rho_j(\mathbf{r})=0}.
\end{eqnarray}

We define the ``orbital densities functional'' $E_{ODF}$ as:
\begin{eqnarray}
\label{ODF8} E_{ODF}[\{\rho_i\}]\equiv
E_{LDA}[\rho] -\frac{1}{2}\sum_i \int d\mathbf{r}
\rho_i(\mathbf{r})\int
d\mathbf{r'}(\rho_i(\mathbf{r'})-\rho^{max}_i(\mathbf{r'}))W(\mathbf{r},\mathbf{r'})\Big|_{\rho_i(\mathbf{r})=0}.
\end{eqnarray}
Please note that the ODF functional [Eq.~(\ref{ODF8})] depends not only
on the total charge density $\rho(\mathbf{r})$ that is a sum of orbital densities 
but also explicitly on a set of orbital densities $\{\rho_i(\mathbf{r})\}$.

Let us investigate the dependence of the functional
[Eq.~(\ref{ODF8})] on the number of electrons $N$. At first please 
note that for the integer values of $N$ corresponding to the integer 
orbital occupancies $n_i$ the value of the functional
[Eq.~(\ref{ODF8})] coincides with the corresponding value of 
the LDA functional. For integer $n_i$ values orbital density 
$\rho_i(\mathbf{r})$ is equal either zero for empty state ($n_i=0$) 
or $\rho^{max}_i(\mathbf{r})$ for occupied state ($n_i=1$). 
In both cases correction term in Eq.~(\ref{ODF8}) vanishes. We will 
show that in accordance with the properties\cite{perdew} of the the
exact density functional $E_{EDF}$ [Eq.~(\ref{LDA23})] this
dependence corresponds to the curve of $E_{ODF}$ versus $N$ as a
series of straight-line segments with derivative discontinuities
at integer values of $N$.

The increase of the total number of electrons $N$ occurs via
consequential increase of the orbital occupancies $n_i$ so that
when $N$ changes from $M$ to $M+1$ the value of the $n_i$
corresponding to the highest occupied orbital for the system with
$M+1$ electrons changes from 0 to 1. The corresponding variation
$\delta\rho(\mathbf{r})$ of the total charge density
$\rho(\mathbf{r})$ will consist exclusively of the variation of
corresponding orbital density $\rho_j(\mathbf{r})$ of particular 
orbital $j$. Then variational derivative of the functional 
[Eq.~(\ref{ODF8})] is equal to:
\begin{eqnarray}
\label{ODF9} \frac{\delta
E_{ODF}}{\delta\rho(\mathbf{r})}=\frac{\delta
E_{ODF}}{\delta\rho_j(\mathbf{r})}= \frac{\delta
E_{LDA}}{\delta\rho(\mathbf{r})}- \int
d\mathbf{r'}(\rho_j(\mathbf{r'})-\frac{1}{2}\rho^{max}_j(\mathbf{r'}))W(\mathbf{r},\mathbf{r'})\Big|_{\rho_j(\mathbf{r})=0}.
\end{eqnarray}
The first term in the right part of Eq.~(\ref{ODF9}) $\delta
E_{LDA}/\delta\rho(\mathbf{r})$ is continuous. However, the
second term depends on the index $j$. When number of electrons $N$
is equal to $M-\omega$ then $j$ corresponds to the highest occupied
orbital for the system with $M$ electrons. However, for
$N=M+\omega$ the index will change to $j+1$ corresponding to the
lowest unoccupied orbital for the system with $M$ electrons or the
highest occupied orbital for the system with $M+1$ electrons. The
value of $\rho_j(\mathbf{r'})$ in the integral in the right part
of Eq.~(\ref{ODF9}) will jump from $\rho^{max}_j(\mathbf{r'})$ to
zero with $N$ going $M-\omega$ to $M+\omega$ for infinitesimally
small $\omega$. That results in the corresponding jump of
variational derivative $\delta
E_{ODF}/\delta\rho(\mathbf{r})$:
\begin{eqnarray}
\label{ODF10} \frac{\delta
E_{ODF}}{\delta\rho(\mathbf{r})}\Big|_{M+\omega}-\frac{\delta
E_{ODF}}{\delta\rho(\mathbf{r})}\Big|_{M-\omega} = \\ \nonumber =
\frac{1}{2}\int
d\mathbf{r'}\Big(\rho^{max}_j(\mathbf{r'})W(\mathbf{r},\mathbf{r'})\Big|_{\rho_j(\mathbf{r})=0}
+\rho^{max}_{j+1}(\mathbf{r'})W(\mathbf{r},\mathbf{r'})\Big|_{\rho_{j+1}(\mathbf{r})=0}\Big).
\end{eqnarray}

In order to show that the variational derivative $\delta
E_{ODF}/\delta\rho(\mathbf{r})$ is constant for number of
electrons varying from $M$ to $M+1$ we need explicit expression
for the variational derivative of LDA functional $\delta
E_{LDA}/\delta\rho(\mathbf{r})$. That could be obtained using
expansion of Eqs.~(\ref{ODF5}) and (\ref{ODF5s}):
\begin{eqnarray}
\label{ODF11}   \frac{\delta
E_{LDA}}{\delta\rho(\mathbf{r})}& = & \frac{\delta
E_{LDA}}{\delta\rho_j(\mathbf{r})}\approx {\frac{\delta
E_{LDA}}{\delta\rho(\mathbf{r})}}\Big|_{\rho_j(\mathbf{r})=0}+
\int d\mathbf{r'}\rho_j(\mathbf{r'})\frac{\delta^2
E_{LDA}}{\delta\rho(\mathbf{r})\delta\rho(\mathbf{r'})}\Big|_{\rho_j(\mathbf{r})=0}  \\ \nonumber& = &
{\Phi(\mathbf{r})}\Big|_{\rho_j(\mathbf{r})=0}+
\int d\mathbf{r'}\rho_j(\mathbf{r'})W(\mathbf{r},\mathbf{r'})\Big|_{\rho_j(\mathbf{r})=0}.
\end{eqnarray}
Then Eq.~(\ref{ODF9}) takes the form:
\begin{eqnarray}
\label{ODF12} \frac{\delta
E_{ODF}}{\delta\rho(\mathbf{r})}& = & \frac{\delta
E_{ODF}}{\delta\rho_j(\mathbf{r})}\approx {\frac{\delta
E_{LDA}}{\delta\rho(\mathbf{r})}}\Big|_{\rho_j(\mathbf{r})=0}+\frac{1}{2}
\int d\mathbf{r'}\rho^{max}_j(\mathbf{r'})\frac{\delta^2
E_{LDA}}{\delta\rho(\mathbf{r})\delta\rho(\mathbf{r'})}\Big|_{\rho_j(\mathbf{r})=0}   \\ \nonumber & = &
\Phi(\mathbf{r})\Big|_{\rho_j(\mathbf{r})=0}+\frac{1}{2}
\int d\mathbf{r'}\rho^{max}_j(\mathbf{r'})W(\mathbf{r},\mathbf{r'})\Big|_{\rho_j(\mathbf{r})=0}.
\end{eqnarray}
The right part of Eq.~(\ref{ODF12}) does not depend on $\rho_j$ and hence it is 
constant for fractional number of electrons $n_j$. 
As derivatives of all terms in LDA functional are continuous the
jump of $\delta E_{ODF}/\delta\rho(\mathbf{r})$ in
Eq.~(\ref{ODF10}) can be assigned to the jump in the variational
derivative of the effective exchange-correlation energy term in
ODF functional $\delta E_{xc}^{ODF}/\delta\rho(\mathbf{r})$. This term can be derived from ODF functional 
[Eq.~(\ref{ODF8})] similarly to the LDA case Eq.~(\ref{LDA44}).

If orbital $\psi_i(\mathbf{r})$ is fixed then variation of orbital density $\rho_i(\mathbf{r})$ 
occurs only via variation of occupancy $n_i$:
\begin{eqnarray}
\label{ODF21s} \delta\rho_i(\mathbf{r}) = \delta n_i|\psi_i(\mathbf{r})|^2.
\end{eqnarray}
Then the second variation of LDA functional is:
\begin{eqnarray}\label{ODF21s2}
\delta^2E_{LDA}& = & \frac{1}{2} \sum_{ij}
\int d\mathbf{r} \delta\rho_i(\mathbf{r})\int
d\mathbf{r'}\delta\rho_j(\mathbf{r'})W(\mathbf{r},\mathbf{r'}) \\ \nonumber & = &
\frac{1}{2} \sum_{ij} \delta n_i\delta n_j\int d\mathbf{r} |\psi_i(\mathbf{r})|^2\int
d\mathbf{r'}|\psi_j(\mathbf{r'})|^2W(\mathbf{r},\mathbf{r'}).
\end{eqnarray}
The second derivative of LDA functional with respect to the occupancy $n_i$ is:
\begin{eqnarray}
\label{ODF21s3}\frac{\delta^2 E_{LDA}}{\delta n_i\delta n_j}
=\int d\mathbf{r} |\psi_i(\mathbf{r})|^2\int
d\mathbf{r'}|\psi_j(\mathbf{r'})|^2W(\mathbf{r},\mathbf{r'}).
\end{eqnarray}

The value of the derivative $\delta^2
E_{LDA}/\delta n_i\delta n_j$ could be
obtained in constrained LDA calculation either in a direct way
or by taking into account that derivative of LDA functional over
orbital occupancy $\delta E_{LDA}/\delta n_i$ is equal to 
one-electron eigenvalue $\epsilon_i$ [see Eqs.~(\ref{LDA62})-(\ref{LDA67})]:
\begin{eqnarray}
\label{ODF23} \frac{\delta^2 E_{LDA}}{\delta n_i\delta n_j}=\frac{\delta
\epsilon_i}{\delta n_j}.
\end{eqnarray}

Using Eqs.~(\ref{ODF21s})-(\ref{ODF23}) ODF functional [Eq.~(\ref{ODF8})]
could be rewritten via occupancies $n_i$:
\begin{eqnarray}
\label{ODF30} E_{ODF}[\rho(\mathbf{r}),\{n_i\}]=
E_{LDA}[\rho(\mathbf{r})] -\frac{1}{2}\sum_i
n_i(n_i-1)\frac{\delta \epsilon_i}{\delta n_i}.
\end{eqnarray}

The derivative of ODF functional [Eq.~(\ref{ODF8})] with respect to the total 
number of electrons $N$ can be expressed via occupancies $n_i$ of partially 
occupied orbital $j$:
\begin{eqnarray}
\label{ODF23s} \frac{\delta E_{ODF}}{\delta N }=\frac{\delta E_{ODF}}{\delta n_j }=
\frac{\delta E_{LDA}}{\delta n_j } + (\frac{1}{2}-n_j)\frac{\delta \epsilon_j}{\delta n_j}=
\epsilon_j+ (\frac{1}{2}-n_j)\frac{\delta \epsilon_j}{\delta n_j}.
\end{eqnarray}
Expanding $\epsilon_j(n_j)$ as a function of $n_j$ and keeping only the first term 
(which is equivalent to keeping only first and second variational derivatives 
in Eq.~(\ref{ODF5})) one can show that derivative $\delta E_{ODF}/\delta N$ 
does not depend on $n_j$:
\begin{eqnarray}
\label{ODF24s} \frac{\delta E_{ODF}}{\delta N } \approx \epsilon_j\Big|_{n_j=0}+
n_j\frac{\delta \epsilon_j}{\delta n_j}\Big|_{n_j=0}+ 
(\frac{1}{2}-n_j)\frac{\delta \epsilon_j}{\delta n_j}\Big|_{n_j=0}=
\epsilon_j\Big|_{n_j=0}+\frac{1}{2}\frac{\delta \epsilon_j}{\delta n_j}\Big|_{n_j=0}.
\end{eqnarray}
Analogue of Eq.~(\ref{ODF10}) demonstrates a jump of $\delta E_{ODF}/\delta N$ 
when number of electrons $N$ is going through an integer value $M$:
\begin{eqnarray}
\label{ODF10s} \frac{\delta
E_{ODF}}{\delta N}\Big|_{M+\omega}-\frac{\delta
E_{ODF}}{\delta N}\Big|_{M-\omega} = (\epsilon_{j+1}-\epsilon_j) 
+\frac{1}{2}\Big(\frac{\delta \epsilon_j}{\delta n_j}+\frac{\delta \epsilon_{j+1}}{\delta n_{j+1}}\Big).
\end{eqnarray}

We have shown that with the accuracy of expansion of
Eq.~(\ref{ODF5}) the variational derivative of the ``orbital
densities functional'' $E_{ODF}$ [Eq.~(\ref{ODF8})] conforms to the conditions for the
{\it{exact}} density functional found in Ref.~\onlinecite{perdew}: it is a constant 
for fractional number of electrons and has a discontinuity when number of electrons 
is going through an integer value. 

Equations (\ref{ODF30}) and (\ref{ODF23s}) are directly related to ``transition state'' 
approach proposed by Slater \cite{slater} to calculate excitation energies. 
In this scheme LDA eigenvalue (Kohn-Sham equations eigenvalue) of the
corresponding one-electron state should be
calculated with its occupancy equal 0.5 (half way between initial and final states 
of excitation process). One can identify a derivative of the ODF functional 
[Eq.~(\ref{ODF30})] over occupancy $n_i$ as a corresponding one-electron 
energy $\epsilon_i^{ODF}$:
\begin{eqnarray}
\label{TS1} \epsilon_i^{ODF}\equiv\frac{\delta E_{ODF}}{\delta n_i }=
\epsilon_i^{LDA}+ (\frac{1}{2}-n_i)\frac{\delta \epsilon_i}{\delta n_i}.
\end{eqnarray} 
If LDA eigenvalue $\epsilon_{i}(n_i)$ is a linear
function of occupancy $n_i$ ($\delta \epsilon_i/\delta n_i=const$) (which is equivalent 
to expansion Eq.~(\ref{ODF10s})) then for an empty state ($n_i=0$):
\begin{eqnarray}
\label{const}
\epsilon_{i}^{LDA}(0.5)& = &
\epsilon_{i}^{LDA}(0)+\frac{1}{2}\frac{\delta \epsilon_i}{\delta n_i}.
\end{eqnarray}
For occupied states ($n_i=1$) a sign plus in Eq.~(\ref{const})
will be replaced by minus. One can see that Eq.~(\ref{TS1}) can reproduce both those cases:
\begin{eqnarray}
\label{TS2} \epsilon_i^{ODF}=\epsilon_i^{LDA}(n_i)+ (\frac{1}{2}-n_i)\frac{\delta \epsilon_i}{\delta n_i}=
\epsilon_{i}^{LDA}(0.5).
\end{eqnarray} 

In Ref.~\onlinecite{GTS} we have introduced an auxiliary functional 
by variation of which Eq.~(\ref{TS2}) giving ``transition state'' 
correction to one-electron energies can be obtained. That functional 
is identical to the functional [Eq.~(\ref{ODF30})]. Keeping in mind 
that we have defined functional [Eq.~(\ref{ODF30})] in order to reproduce 
properties of the exact density functional\cite{perdew} one can 
arrive to conclusion that eigenvalues calculated in ``transition state'' 
method are a good approximation to the eigenvalues of the {\it{exact}} functional.

The minimization of the functional Eq.~(\ref{ODF8}) will give a set
of orbital densities $\{\rho_i(\mathbf{r})\}$ and hence the total
charge density $\rho(\mathbf{r})$ corresponding to the ground
state of the system. It is useful to derive equations for
{\it{fluctuations}} of the orbital densities
$\{\delta\rho_i(\mathbf{r})\}$ around the average ground state
functions $\{\rho_i(\mathbf{r})\}$. This ``ground state'' charge density 
$\rho(\mathbf{r})$ corresponds to the minimum of ODF functional [Eq.~(\ref{ODF8})] 
but not of the LDA functional so that one must use in the calculations constrain 
potential $\Phi(\mathbf{r})$ (see Appendix \ref{constrain}, Eq.~(\ref{LDA61})). 
Analogous to the expansion Eqs.~(\ref{ODF5}) and (\ref{ODF5s}) one can write 
for the LDA part of ODF functional:
\begin{eqnarray}
\label{ODF13}
E_{LDA}[\{\rho_i+\delta\rho_i\}] \approx
E_{LDA}[\{\rho_i\}]+ \sum_i\int d\mathbf{r}
\delta\rho_i(\mathbf{r}){\Phi(\mathbf{r})}
\\ \nonumber  + \frac{1}{2}\sum_i\sum_j\int d\mathbf{r} \delta\rho_i(\mathbf{r})\int
d\mathbf{r'}\delta\rho_j(\mathbf{r'})W(\mathbf{r},\mathbf{r'}).
\end{eqnarray}
For the correction term Eq.~(\ref{ODF6}) the corresponding
expression is:
\begin{eqnarray}
\label{ODF14}
E_{corr}[\{\rho_i+\delta\rho_i\}] \approx
E_{corr}[\{\rho_i\}]-\\ \nonumber - \sum_i\int
d\mathbf{r} \delta\rho_i(\mathbf{r})\int
d\mathbf{r'}(\rho_i(\mathbf{r'})-\frac{1}{2}\rho^{max}_i(\mathbf{r'}))W(\mathbf{r},\mathbf{r'})
\\ \nonumber  - \frac{1}{2}\sum_i\int d\mathbf{r} \delta\rho_i(\mathbf{r})\int
d\mathbf{r'}\delta\rho_i(\mathbf{r'})W(\mathbf{r},\mathbf{r'}).
\end{eqnarray}
Both Eqs.~(\ref{ODF13}) and (\ref{ODF14}) become exact if one
supposes that $\frac{\delta^2
E_{LDA}}{\delta\rho(\mathbf{r})\delta\rho(\mathbf{r'})}=W(\mathbf{r},\mathbf{r'})$ does
not depend on fluctuations $\delta\rho_i(\mathbf{r})$ so that all
variational derivatives higher than the second order are equal to
zero. In the following we assume that this approximation is valid.
Then for the total ODF functional one has:
\begin{eqnarray}
\label{ODF15}
E_{ODF}[\{\rho_i+\delta\rho_i\}] =
E_{ODF}[\{\rho_i\}]+ \\ \nonumber + \sum_i\int
d\mathbf{r} \delta\rho_i(\mathbf{r})\Big(\Phi(\mathbf{r})- \int
d\mathbf{r'}(\rho_i(\mathbf{r'})-\frac{1}{2}\rho^{max}_i(\mathbf{r'}))W(\mathbf{r},\mathbf{r'}) \Big)
\\ \nonumber  + \frac{1}{2}\sum_i\sum_{j\neq i}\int d\mathbf{r} \delta\rho_i(\mathbf{r})\int
d\mathbf{r'}\delta\rho_j(\mathbf{r'})W(\mathbf{r},\mathbf{r'}).
\end{eqnarray}

It is convenient to introduce effective ODF potential
$V^{ODF}_i(\mathbf{r})$:
\begin{eqnarray}
\label{ODF16} V^{ODF}_i(\mathbf{r})& \equiv & \Phi(\mathbf{r})- \int
d\mathbf{r'}(\rho_i(\mathbf{r'})-\frac{1}{2}\rho^{max}_i(\mathbf{r'}))W(\mathbf{r},\mathbf{r'}).
\end{eqnarray}

Using Eqs.~(\ref{ODF15}) and (\ref{ODF16}) one can define a Hamiltonian
for the density matrix fluctuation operators
$\widehat{\delta\rho}_i(\mathbf{r})\equiv
\widehat{\rho}_i(\mathbf{r})-\rho_i(\mathbf{r})$ (here ground
state orbital density can be considered as an average value of
density matrix operator
$\rho_i(\mathbf{r})=\langle\widehat{\rho}_i(\mathbf{r})\rangle$):
\begin{eqnarray}
\label{ODF17} \widehat{H} \equiv \sum_i\int d\mathbf{r}
\widehat{\delta\rho}_i(\mathbf{r})V^{ODF}_i(\mathbf{r})+
\frac{1}{2}\sum_i\sum_{j\neq i}\int d\mathbf{r}
\widehat{\delta\rho}_i(\mathbf{r})\int
d\mathbf{r'}\widehat{\delta\rho}_j(\mathbf{r'})W(\mathbf{r},\mathbf{r'}).
\end{eqnarray}

The orbital density matrix operators
$\widehat{\rho}_i(\mathbf{r})$ could be expressed via orbital
occupancy operators $\widehat{n}_i$ (if orbitals $\psi_i(\mathbf{r})$ are fixed 
and the variation of orbital density $\rho_i(\mathbf{r})$ occurs only via variation 
of occupancies $n_i$):
\begin{eqnarray}
\label{ODF20}
\widehat{\rho}_i(\mathbf{r})=\widehat{n}_i|\psi_i(\mathbf{r})|^2.
\end{eqnarray}
The fluctuation Hamiltonian Eq.~(\ref{ODF17}) can be expressed using Eqs.~(\ref{ODF21s})-(\ref{ODF23}) 
via occupancy operators $\widehat{n}_i$ and 
their average values $n_i=\langle\widehat{n}_i\rangle$ ($\widehat{\delta n}_i\equiv \widehat{n}_i-\langle\widehat{n}_i\rangle$)
\begin{eqnarray}
\label{H_fluct} \widehat{H} & = & 
\sum_i(\epsilon_{i}+(\frac{1}{2}-n_i)\frac{\partial
\epsilon_{i}}{\partial n_i})
\widehat{\delta n}_i+\frac{1}{2}\sum_{i}\sum_{j\neq i} \frac{\partial
\epsilon_{i}}{\partial n_j} \widehat{\delta n}_i
\widehat{\delta n}_j.
\end{eqnarray}

The orbital densities $\rho_i(\mathbf{r})$ are defined by orbitals
$\psi_i(\mathbf{r})$ in expression for charge density in
Eq.~(\ref{ODF1}). In the density functional theory it is usually
assumed that those orbitals are solutions of Kohn-Sham equations 
[Eqs.~(\ref{LDA46}) and (\ref{LDA54})]
and so minimize functional Eq.~(\ref{LDA44}). However, any unitary
transformation (defined by unitary matrix $U$) of the set of the
functions $\psi_i(\mathbf{r})$ produces a new set of orbitals:
\begin{eqnarray} 
\label{ODF27}
\widetilde{\psi}_i(\mathbf{r})\equiv\sum_j U_{ij}\psi_j(\mathbf{r})
\end{eqnarray}
corresponding to the same charge density $\rho(\mathbf{r})$. This
new set of orbitals can be used to define ODF functional in
Eq.~(\ref{ODF8}) an so this functional is orbital-dependent.

In order to remove uncertainty in choosing the orbital set one
needs to impose an additional condition. We propose the following
way to do it. The Hamiltonian Eq.~(\ref{ODF17}) can be used to
calculate the contribution to the energy from fluctuations:
\begin{eqnarray}
\label{ODF18} \langle\widehat{H}\rangle = \frac{1}{2}\sum_i\sum_{j\neq i}\int
d\mathbf{r} \int d\mathbf{r'}\langle\widehat{\delta\rho}_i(\mathbf{r})
\widehat{\delta\rho}_j(\mathbf{r'})\rangle W(\mathbf{r},\mathbf{r'}).
\end{eqnarray}
(The average value of fluctuations
$\langle\widehat{\delta\rho}_i(\mathbf{r})\rangle$ is equal to zero so the first 
term in Eq.~(\ref{ODF17}) 
does not give contribution to the fluctuation energy.) Taking
into account Eqs.~(\ref{ODF13}) and (\ref{ODF14}) one can separate
contributions to Eq.~(\ref{ODF18}) from LDA functional where
summation over $i$ and $j$ is performed including terms with
$i=j$ and a correction term Eq.~(\ref{ODF14}). 

The LDA part is:
\begin{eqnarray}
\label{ODF18s} \langle\widehat{H}_{LDA}\rangle & = & \frac{1}{2}\sum_i\sum_{j}\int
d\mathbf{r} \int d\mathbf{r'}\langle\widehat{\delta\rho}_i(\mathbf{r})
\widehat{\delta\rho}_j(\mathbf{r'})\rangle W(\mathbf{r},\mathbf{r'}) \\ \nonumber 
& = & \frac{1}{2}\int
d\mathbf{r} \int d\mathbf{r'}\langle\sum_i\widehat{\delta\rho}_i(\mathbf{r})
\sum_{j}\widehat{\delta\rho}_j(\mathbf{r'})\rangle W(\mathbf{r},\mathbf{r'}) \\ \nonumber
& = & \frac{1}{2}\int
d\mathbf{r} \int d\mathbf{r'}\langle\widehat{\delta\rho}(\mathbf{r})
\widehat{\delta\rho}(\mathbf{r'})\rangle W(\mathbf{r},\mathbf{r'}).
\end{eqnarray}

Correction term gives a
negative contribution equal to:
\begin{eqnarray}
\label{ODF19} \langle\widehat{H}_{corr}\rangle = -\frac{1}{2}\sum_i\int
d\mathbf{r} \int d\mathbf{r'}\langle\widehat{\delta\rho}_i(\mathbf{r})
\widehat{\delta\rho}_i(\mathbf{r'})\rangle W(\mathbf{r},\mathbf{r'}).
\end{eqnarray}

From Eq.~(\ref{ODF18s}) one can see that LDA contribution to fluctuation energy 
is defined by the total charge density fluctuations $\langle\widehat{\delta\rho}(\mathbf{r})
\widehat{\delta\rho}(\mathbf{r'})\rangle$ and so does not depend on the orbitals definition. 
Then the minimum of the fluctuation energy Eq.~(\ref{ODF18}) is achieved
when the absolute value of correction contribution
Eq.~(\ref{ODF19}) has a maximum.

The correction term contribution in Eq.~(\ref{ODF19}) can be calculated using 
expression of fluctuations Hamiltonian via fluctuation occupancy operators 
$\widehat{\delta n}_i$ [Eq.~(\ref{H_fluct})]:
\begin{eqnarray}
\label{ODF22} \langle\widehat{H}_{corr}\rangle =
-\frac{1}{2}\sum_i\langle\widehat{\delta n}_i \widehat{\delta n}_i\rangle\frac{\partial\epsilon_{i}}{\partial n_i}.
\end{eqnarray}

As the average value of square of occupancy fluctuations
$\langle\widehat{\delta n}_i \widehat{\delta n}_i\rangle$ depends on the
specific properties of the system, the only way to minimize fluctuation energy 
$\langle\widehat{H}\rangle$ [Eq.~(\ref{ODF18})] is to maximize $\partial\epsilon_{i}/\partial n_i$. 
Using Eqs.~(\ref{ODF21s3}) and (\ref{ODF23}) this parameter can be expressed via orbitals as:
\begin{eqnarray}
\label{ODF21s4}
\frac{\partial\epsilon_{i}}{\partial n_i}=\frac{\delta^2 E_{LDA}}{\delta n_i^2}
=\int d\mathbf{r} |\psi_i(\mathbf{r})|^2\int
d\mathbf{r'}|\psi_i(\mathbf{r'})|^2W(\mathbf{r},\mathbf{r'}).
\end{eqnarray}

For new set of the functions $\widetilde{\psi}_i(\mathbf{r})$ 
[Eq.~(\ref{ODF27})] one has:
\begin{eqnarray}
\label{ODF28} \frac{\delta \tilde{\epsilon}_i}{\delta \tilde{n}_i}
=\sum_{jj'll'} U_{ij}U_{ij'}^*U_{il}U_{il'}^*\int d\mathbf{r}\int d\mathbf{r'}
\psi_j(\mathbf{r})\psi_{j'}^*(\mathbf{r})W(\mathbf{r},\mathbf{r'})\psi_{l}(\mathbf{r'})\psi_{l'}^*(\mathbf{r'}).
\end{eqnarray}
Using Eq.~(\ref{ODF28}) and the fact that the derivatives $\delta \epsilon_i/\delta n_i$ 
are always positive one can define a functional of unitary matrix $U$:
\begin{eqnarray}
\label{ODF29} F[U]\equiv\sum_i\frac{\delta
\tilde{\epsilon}_i}{\delta \tilde{n}_i} = 
\sum_i
\sum_{jj'll'} U_{ij}U_{ij'}^*U_{il}U_{il'}^*\int d\mathbf{r}\int d\mathbf{r'}
\psi_j(\mathbf{r})\psi_{j'}^*(\mathbf{r})W(\mathbf{r},\mathbf{r'})\psi_{l}(\mathbf{r'})\psi_{l'}^*(\mathbf{r'})
\end{eqnarray}
maximization of which one can use as a condition to determine the
matrix $U$ and hence the optimal set of orbitals
$\psi_i(\mathbf{r})$ to define ODF functional.

The function $W(\mathbf{r},\mathbf{r'})$ [Eq.~(\ref{LDA60})] is defined 
as screened effective interaction between density fluctuations 
$\delta\rho(\mathbf{r}),\delta\rho(\mathbf{r'})$ and so should decay 
with increasing of $|\mathbf{r}-\mathbf{r'}|$ value. Then the more localized 
in space orbitals $\psi_i(\mathbf{r})$ are, the larger should be 
the integral Eq.~(\ref{ODF21s4}) value. 
One of possible choices for these orbitals could be Wannier functions. 
Marzari and Vanderbilt 
in Ref.~\onlinecite{vanderbildt} proposed the condition
of maximum localization to determine the procedure to calculate Wannier
functions for multi-band case (two orthonormal sets of functions, 
Wannier and Bloch, are connected via unitary transformation 
so Wannier functions can be considered as a particular choice 
of unitary matrix $U$ in Eq.~(\ref{ODF27})). The requirement 
of maximum localization should lead to the reasonably maximized
values of parameter $\partial\epsilon_{i}/\partial n_i$ in
Eq.~(\ref{ODF21s4}) and hence to minimization the fluctuations energy 
Eq.~(\ref{ODF18}). Then Wannier functions obtained via procedure 
proposed in Ref.~\onlinecite{vanderbildt} are a good choice 
for a set of orbitals to define ODF functional. 

In the following we will assume that orbital densities 
are defined not by one-electron eigenfunctions 
$\psi_i(\bf{r})$ as in Eq.~(\ref{ODF1}) but by 
Wannier functions $W_n({\bf r})$ and their occupancies 
$Q_{n}$ calculated via procedure described in Sec. \ref{W_def}:
\begin{eqnarray}
\label{ODF31} 
\rho(\mathbf{r}) = \sum_n \rho_n(\bf{r}) ,\\ \nonumber
\rho_n(\mathbf{r})=Q_{n}|W_n({\bf r})|^2.
\end{eqnarray}

Using Eqs.~(\ref{ODF30})-(\ref{ODF20}) ODF functional 
[Eq.~(\ref{ODF8})] could be rewritten with Wannier functions 
occupancy operators $\widehat{Q}_n$ [Eq.~(\ref{Q_WF2})]
and their average values $Q_{n}=\langle\widehat{Q}_n\rangle$.
The corresponding functional will be
analogous to Eq.~(\ref{ODF30}) but with occupancies
$Q_{n}$ [Eq.~(\ref{Q_WF})] and energies $E_{n}$ [Eq.~(\ref{E_WF})] 
corresponding to Wannier functions:
\begin{eqnarray}
\label{functional-GTS} E_{ODF}= E_{LDA}-\frac{1}{2} \sum_n
\frac{\partial E_{n}}{\partial
Q_{n}}Q_{n}(Q_{n}-1).
\end{eqnarray}

The variation of the functional Eq.~(\ref{functional-GTS}) 
will produce (see Appendix \ref{project}) one-electron Hamiltonian
$\widehat{H}^0_{ODF}$ in the form of projection operator:
\begin{eqnarray}
\label{H_corr}
\widehat{H}^0_{ODF}  = \widehat{H}_{LDA}+\sum_{n} \delta V_{n} \widehat{Q}_n  
=  \widehat{H}_{LDA}+\sum_{n\bf T} |W_{n}^{\bf
T} \rangle \delta V_{n} \langle W_{n}^{\bf T}|,
\end{eqnarray}
$|W_{n}^{\bf T} \rangle$ in Eq.~(\ref{H_corr}) are Wannier
functions [Eq.~(\ref{W-b})] and $\delta V_{n}$ are:
\begin{eqnarray}
\label{dV-GTS} \delta V_{n} = \frac{\partial E_{n}}{\partial
Q_{n}}(\frac{1}{2}-Q_{n}).
\end{eqnarray}
The values of derivatives $\partial E_{n}/\partial Q_{n}$ 
should be determined in constrained LDA calculations (see Sec.~\ref{W_def}).

Equations~(\ref{functional-GTS})-(\ref{dV-GTS}) are identical 
to the equations of ``generalized transition state'' (GTS) method that 
we have developed in Ref.~\onlinecite{GTS}. 
The basis of GTS method was an idea that one-electron energies 
corresponding to Wannier functions should have a meaning 
of the removal (addition) energies for electrons from (to) the corresponding states. 
That was realized by using ``transition state'' scheme \cite{slater} generalized 
to Wannier functions instead of eigenfunctions. It is remarkable that 
the same equations can be obtained by introducing correction to the LDA functional 
restoring the properties of the {\it{exact}} density functional theory.~\cite{perdew}

The potential correction operator [Eqs.~(\ref{H_corr}) and (\ref{dV-GTS})] shifts 
the energies of Wannier functions on $\delta V_{n}$ values which are negative 
($\delta V_{n}=-\frac{1}{2}\frac{\partial E_{n}}{\partial Q_{n}}$) 
for the occupied states ($Q_{n}=1$) and positive 
($\delta V_{n}=\frac{1}{2}\frac{\partial E_{n}}{\partial Q_{n}}$) 
for the empty states ($Q_{n}=0$). That gives valence bands pushed down 
and conduction bands pushed up comparing with standard LDA results 
increasing energy separation between them (energy gap value) which 
is systematically underestimated in LDA. It was shown in Ref.~\onlinecite{GTS} 
that calculations with potential correction 
[Eqs.~(\ref{H_corr}) and (\ref{dV-GTS})] result in a much better agreement 
with experimental energy gap values for semiconductor (Si), band insulator 
(MgO), Mott insulator (NiO), and Peierls insulator BaBiO$_3$. 

The Hamiltonian $\widehat{H}^0_{ODF}$ [Eq.~(\ref{H_corr})]
can be considered as a static mean-field approximation to a general problem. 
Wannier functions analogue of the fluctuation Hamiltonian 
Eq.~(\ref{H_fluct}) is:
\begin{eqnarray}
\label{H_corr3} \widehat{H}_{ODF} & = & \widehat{H}^0_{ODF}
+\frac{1}{2}\sum_{n}\sum_{n'\neq n}\frac{\partial
E_{n}}{\partial
Q_{n'}}(\widehat{Q}_n-\langle\widehat{Q}_n\rangle)(\widehat{Q}_{n'}-\langle\widehat{Q}_{n'}\rangle).
\end{eqnarray}

The first and second parts of Hamiltonian Eq.~(\ref{H_corr3}) are
not~(!)~{\it{noninteracting}} Hamiltonian and {\it{interaction}} term
as it is usually defined in model Hubbard and Anderson
Hamiltonians. The first part $\widehat{H}^{0}_{ODF}$ is equivalent
to the Hartree-Fock approximation Hamiltonian determined by the
average values of Wannier functions occupancies $\langle\widehat{Q}_n\rangle$
and the second part describes interaction between fluctuations
around $\langle\widehat{Q}_n\rangle$. As these average values are determined from
the solution of the full Hamiltonian Eq.~(\ref{H_corr3}), that
defines a self-consistent calculation scheme. In contrast to LDA+$U$\cite{ani91,ani97}
and LDA+DMFT \cite{LDADMFT} methods there is no ``double counting'' problem in this
Hamiltonian because there were no ``merging'' of LDA and Hubbard
model concepts here and both terms in Eq.~(\ref{H_corr3}) were
derived from the same functional Eq.~(\ref{functional-GTS}).

The problem defined by the Hamiltonian Eq.~(\ref{H_corr3}) can be
solved by any of the methods developed to treat many-body effects.
In the present work we have used dynamical mean-field theory
(DMFT)\cite{vollha93,pruschke,georges96} (see Sec. \ref{WF_DMFT}).

\section{Definition and construction of Wannier functions}
\label{W_def}
The orbital projection calculation scheme for Wannier functions
(WFs) used in the present work was described in details in the
earlier paper\cite{WF-DMFT} where the LDA+DMFT (DMFT -- dynamical
mean-field theory) method in the Wannier function basis set was
proposed. Below we present the main formulas of this scheme.

The concept of WFs has a very important place in the electron
theory in solids since its first introduction in 1937 by Wannier.~\cite{wannier} 
WFs are the Fourier transformation of Bloch
states $|\psi_{i\bf k}\rangle$,
\begin{eqnarray}
\label{WF_psi_def} |W_{i}^{\bf T}\rangle& = &
\frac{1}{\sqrt{N}}\sum_{\bf k} e^{-i{\bf kT}}|\psi_{i{\bf
k}}\rangle,
\end{eqnarray}
where $N$ is the number of discrete ${\bf k}$ points in the first
Brillouin zone (or, the number of cells in the crystal) and {\bf
T} is lattice translation vector.

Wannier functions are not uniquely defined for a many-band case
because for a certain set of bands any orthogonal linear
combination of Bloch functions $|\psi_{i\bf k}\rangle$ can be used
in Eq.~(\ref{WF_psi_def}). In general it means the freedom to choose 
the of unitary transformation matrix $U^{({\bf k})}_{ji}$ for Bloch
functions:~\cite{vanderbildt}
\begin{eqnarray}
\label{psi_def} |\widetilde\psi_{i\bf k}\rangle & = & \sum_j
U^{({\bf k})}_{ji} |\psi_{j\bf k}\rangle.
\end{eqnarray}
There is no rigorous way to define $U^{({\bf k})}_{ji}$. This
calls for an additional restriction on the properties of WFs.
Among others, Marzari and Vanderbilt~\cite{vanderbildt} used the
condition of maximum localization for WFs, resulting in a
variational procedure to calculate $U^{({\bf k})}_{ji}$. To get a
good initial guess the authors of Ref.~\onlinecite{vanderbildt} proposed
choosing a set of localized trial orbitals $|\phi_n\rangle$ and
projecting them onto the Bloch functions $|\psi_{i\bf k}\rangle$.
It was found that this starting guess is usually quite good. This
fact later led to the simplified calculating scheme 
in Ref.~\onlinecite{pickett} where the variational procedure was abandoned 
and the result of the aforementioned projection was considered 
as the final step.

For the projection procedure used in the present work one needs to
identify the set of bands and corresponding set of localized trial
orbitals $|\phi_n\rangle$. The choice of bands and
orbitals is determined by the physics of the system under
consideration.

The set of bands can be defined either by the band indices of the
corresponding Bloch functions ($N_1,...,N_2$), or by choosing the
energy interval ($E_1,E_2$) in which the bands are located.
Nonorthogonalized WFs in reciprocal space $|\widetilde{W}_{n\bf
k}\rangle$ are then the projection of the set of site-centered
atomiclike trial orbitals $|\phi_n\rangle$ on the Bloch functions
$|\psi_{i\bf k}\rangle$ of th chosen bands defined by band
indices ($N_1$ to $N_2$) or by energy interval ($E_1,E_2$):
\begin{eqnarray}
\label{WF_psi} |\widetilde{W}_{n\bf k}\rangle & \equiv &
\sum_{i=N_1}^{N_2} |\psi_{i\bf k}\rangle\langle\psi_{i\bf
k}|\phi_n\rangle = \sum_{i(E_1\le \varepsilon_{i}({\bf k})\le
E_2)} |\psi_{i\bf k}\rangle\langle\psi_{i\bf k}|\phi_n\rangle.
\end{eqnarray}

In the present work we have used LMTO method\cite{LMTO} to solve
a band structure problem and the trial orbitals $|\phi_n\rangle$
were LMTOs. The coefficients
$\langle\psi_{i\bf k}|\phi_n\rangle$ in Eq.~(\ref{WF_psi}) define
(after orthonormalization) the unitary transformation matrix
$U^{({\bf k})}_{ji}$ in Eq.~(\ref{psi_def}).

In any DFT method the Kohn-Sham orbitals are expanded through
the certain basis:
\begin{eqnarray}\label{psi}
|\psi_{i\bf k}\rangle & = & \sum_{\mu} c^{\bf k}_{\mu
i}|\phi_{\mu}^{\bf k}\rangle.
\end{eqnarray}
The basis functions of the LMTO method are Bloch sums 
of the site-centered orbitals:
\begin{eqnarray} \label{psik}
\phi_{\mu}^{\bf k}({\bf r}) & = & \frac{1}{\sqrt{N}}
\sum_{\bf T} e^{i\bf kT} \phi_{\mu}({\bf r}-{\bf R}_q-{\bf T}),
\end{eqnarray}
where $\mu$ is the combined index representing $qlm$ ($q$ is the
atomic number index in the unit cell, $lm$ are orbital and magnetic
quantum numbers), ${\bf R}_q$ is the position of atom in the
unit cell.

For the {\it orthogonal} LMTO basis $c^{\bf k}_{\mu
i}=\langle\phi_{\mu}|\psi_{i \bf k}\rangle $ and hence

\begin{eqnarray}
\label{WF} |\widetilde{W}_{n\bf k}\rangle & = &
\sum_{i=N_1}^{N_2} |\psi_{i\bf k}\rangle c_{ni}^{{\bf k}*}
 =  \sum_{i=N_1}^{N_2} \sum_{\mu} c^{\bf k}_{\mu i} c_{ni}^{{\bf k}*}
|\phi_{\mu}^{\bf k}\rangle  = \sum_{\mu} \tilde{b}^{\bf k}_{\mu n}
|\phi_{\mu}^{\bf k}\rangle,
\end{eqnarray}
with
\begin{equation}
\label{coef_b} \tilde{b}^{\bf k}_{\mu n} \equiv \sum_{i=N_1}^{N_2}
c^{\bf k}_{\mu i} c_{ni}^{{\bf k}*}.
\end{equation}
For a nonorthogonal basis set orthogonalization of the Hamiltonian
must be done before using Eq.~(\ref{coef_b}).

In order to orthonormalize the WFs [Eq.~(\ref{WF})] one needs to
calculate the overlap matrix $O_{nn'}({\bf k})$, 
\begin{eqnarray}
\label{O-S} O_{nn'}({\bf k})&\equiv& \langle\widetilde{W}_{n\bf
k}|\widetilde{W}_{n'\bf k}\rangle = \sum_{i=N_1}^{N_2} c^{\bf
k}_{ni} c_{n'i}^{{\bf k}*},
\end{eqnarray}
and its inverse square root $S_{nn'}({\bf k})$,
\begin{eqnarray}
\label{SS} S_{nn'}({\bf k}) &\equiv& O^{-1/2}_{nn'}({\bf k}).
\end{eqnarray}

From Eqs.~(\ref{WF}) and (\ref{SS}) the orthonormalized WFs in
${\bf k}$ space $|W_{n\bf k}\rangle$ can be obtained as
\begin{eqnarray}
\label{WF_orth} |W_{n\bf k}\rangle = \sum_{n'} S_{nn'}({\bf k})
|\widetilde{W}_{n'\bf k}\rangle =\sum_{i=N_1}^{N_2} |\psi_{i\bf
k}\rangle \bar{c}_{ni}^{{\bf k}*}  = \sum_{\mu} b^{\bf k}_{\mu n}
|\phi_{\mu} ^{\bf k}\rangle,
\end{eqnarray}
with
\begin{eqnarray}
\label{coef_WF_Bloch} \bar{c}_{ni}^{{\bf k}*}\equiv
\langle\psi_{i\bf k}|W_{n\bf k}\rangle =\sum_{n'} S_{nn'}({\bf k})
c_{n'i}^{{\bf k}*},
\end{eqnarray}
\begin{eqnarray}
\label{coef_WF_LMTO} b^{\bf k}_{\mu n} \equiv
\langle\phi_{\mu}^{\bf k}|W_{n\bf k}\rangle= \sum_{i=N_1}^{N_2}
c^{\bf k}_{\mu i} \bar{c}_{ni}^{{\bf k}*}.
\end{eqnarray}

Using Eqs.~(\ref{WF_orth})-(\ref{coef_WF_LMTO}) one can find energies of WFs:

\begin{eqnarray} \label{E_WF}
E_{n}& = & \langle W_{n}^{\bf T}|
  \widehat{H}|W_{n}^{\bf T}\rangle
 =  \langle W_{n}^{\bf T}|
  \biggl(\sum_{i,\bf k} |\psi_{i\bf k}\rangle
    \epsilon_{i}({\bf k})\langle\psi_{i\bf k}|
  \biggr)|W_{n}^{\bf T}\rangle  \nonumber \\
           & = & \frac{1}{N}
       \sum_{\bf k}\sum_{i=N_1}^{N_2} \bar{c}^{\bf k}_{ni}
      \bar{c}_{n'i}^{{\bf k}*}\epsilon_{i}({\bf k})
\end{eqnarray}
and their occupancies:
\begin{eqnarray} \label{Q_WF}
Q_{n} & = & \langle W_{n}^{\bf T}|
  \biggl(\sum_{i,\bf k} |\psi_{i\bf k}\rangle
    \theta(E_f - \epsilon_{i}({\bf k}) )\langle\psi_{i\bf k}|
  \biggr)|W_{n}^{\bf T}\rangle  \nonumber \\
           & = & \frac{1}{N}
       \sum_{\bf k}\sum_{i=N_1}^{N_2} \bar{c}^{\bf k}_{ni}
      \bar{c}_{n'i}^{{\bf k}*}\theta(E_f - \epsilon_{i}({\bf k}) ),
\end{eqnarray}
where $\epsilon_{i}({\bf k})$ is the eigenvalue for a particular band,
$\theta(x)$ is the step function, $E_f$ is the Fermi energy.

The transformation from LMTO to the WF basis set is defined by the
explicit form of WFs Eqs.~(\ref{WF_orth}) and (\ref{coef_WF_LMTO}),
and by the expressions for matrix elements of the Hamiltonian and
density matrix operators in WF basis [Eqs.~(\ref{E_WF}) and
(\ref{Q_WF})]. The transformation from WF to LMTO basis can also
be defined using Eq.~(\ref{WF_orth}). Such transformation is
necessary in calculations using ODF correction potential in the form of projection operator
Eq.~(\ref{H_corr}). It will also be used for constrained LDA calculations needed
to determine the values of WFs energy derivatives with respect to their occupancies 
$\partial E_{n}/\partial Q_{n'}$ which define potential correction parameters 
$\delta V_{n}$ in Eqs.~(\ref{H_corr}) and (\ref{dV-GTS}) and effective Coulomb parameters 
$U_{nn'}$ in ODF fluctuation Hamiltonian in the Hubbard form 
Eqs.~(\ref{H_corr4})-(\ref{H_corr6}).
For example if constrain
potential operator is diagonal in WF basis ($H^{constr}_{nn'}=\Lambda
_n\delta_{nn'}$), then in LMTO basis its matrix elements can be
calculated via:
\begin{eqnarray} \label{H_constrain}
\widehat{H}^{constr}  =  \sum_{n,{\bf T}} |W_{n}^{\bf T} \rangle
\Lambda_n \langle W_{n}^{\bf T}|,
\end{eqnarray}
\begin{equation} \label{W-b}
|W_{n}^{\bf T} \rangle  =  \sum_{j,{\bf k}} e^{-i{\bf kT}} b^{\bf
k}_{jn} |\phi_j^{\bf k}\rangle ,
\end{equation}
\begin{eqnarray} 
\label{H_constrain2}
H^{constr}_{\mu\nu}(\mathbf{k})  =  \langle \phi_{\mu}^{\mathbf{k}}
|\widehat{H}^{constr}| \phi_{\nu}^{\mathbf{k}}\rangle  =  \sum_n
b_{\mu n}^{\mathbf{k}} \Lambda_n b_{\nu n}^{\mathbf{k}*}.
\end{eqnarray}

\section{Wannier functions in the Green-function formalism}
\label{WF_GF}

In many-body theory the system is usually described not by Bloch
functions $|\psi_{i\bf k}\rangle$ [Eq.~(\ref{psi})] and their energies
$\epsilon_{i}({\bf k})$ but by the Green function
\begin{eqnarray}
\label{Green1} G({\bf r},{\bf r'},\varepsilon) & = & \frac{1}{N}
\sum_{\bf k} G^{\bf
k}({\bf r},{\bf r'},\varepsilon)= \frac{1}{N}
\sum_{\bf k} \sum_{\mu \nu}
\phi_{\mu}^{\bf k}({\bf r}) G^{\bf k}_{\mu \nu}(\varepsilon)
\phi^{*\bf k}_{\nu}({\bf r'}).
\end{eqnarray}
The matrix Green function $G^{\bf k}_{\mu \nu}(\varepsilon)$ is
defined via the noninteracting Hamiltonian matrix $H^0_{\mu
\nu}({\bf k})$ [Eq.~(\ref{H_corr6}) and (\ref{H_constrain2})] 
and the matrix self-energy $\Sigma^{\bf k}_{\mu
\nu}(\varepsilon)$ [Eq.~(\ref{Sigma_full})] as
\begin{eqnarray}
\label{Green2} G^{\bf k}_{\mu \nu}(\varepsilon) =  [\varepsilon -
\widehat{H}^0({\bf k}) - \widehat{\Sigma}(\varepsilon,{\bf
k})+i\eta]^{-1}_{\mu \nu}.
\end{eqnarray}
We define the nonorthonormalized WF obtained by projecting the trial
orbital $\phi_n({\bf r})$ on the Hilbert subspace determined by the
Green function [Eq.~(\ref{Green1})] in the energy interval ($E_1,E_2$),
namely,
\begin{eqnarray}
\label{WF1} \widetilde{W}_{n{\bf k}}({\bf r}) & = & -\frac{1}{\pi}
{\rm Im} \int_{E_1}^{E_2} d\varepsilon \int d{\bf r}' G^{\bf k}({\bf
r},{\bf r}',\varepsilon) \phi_n^{\bf k}({\bf r}')
=  \sum_{\mu } \tilde{b}_{\mu n}^{\bf k} \phi_{\mu}^{\bf k}({\bf r})\\
\nonumber
\end{eqnarray}
and
\begin{eqnarray}
\label{coef_green_b}
\tilde{b}_{\mu n}^{\bf k} & \equiv & -\frac{1}{\pi} {\rm Im}
\int_{E_1}^{E_2} d\varepsilon  G^{\bf k}_{\mu n}(\varepsilon).
\end{eqnarray}

In the noninteracting case, the self-energy
$\hat\Sigma(\varepsilon,{\bf k})$ is absent and hence we have
\begin{eqnarray}
\label{Green3} G^{\bf k}_{\mu \nu}(\varepsilon)
  &=&\sum_i \frac{c^{\bf k}_{\mu i} c_{\nu i}^{{\bf k}*}}
{\varepsilon - \epsilon_{i}({\bf k})+i\eta},
\end{eqnarray}
where $c^{\bf k}_{\mu i}$ are the eigenvectors, and
$\epsilon_{i}({\bf k})$ are the eigenvalues of $\widehat{H}^0({\bf k})$.
Thus $\tilde{b}_{\mu n}^{\bf k}$ in Eq.~(\ref{coef_green_b}) becomes
\begin{eqnarray}
\label{b_koef}  \tilde{b}_{\mu n}^{\bf k} =
\sum_{i=N_1}^{N_2}  c^{\bf k}_{\mu i} c_{ni}^{{\bf k}*},
\end{eqnarray}
where $N_1,N_2$ are the band numbers corresponding to the energy interval
($E_1,E_2$). Since this recovers the result of Eq.~(\ref{coef_b}),
we demonstrated that our general definition of WFs [Eq.~(\ref{WF1})] via Green
function reduces to that in terms of Bloch functions [Eq.~(\ref{WF})] in
Sec.~\ref{W_def}.

To orthonormalize $\widetilde{W}_{n{\bf k}}({\bf r})$ defined in Eq.~(\ref{WF1}),
one can just follow the orthonormalizing procedure made in Sec.~\ref{W_def}
[Eqs.~(\ref{O-S})-(\ref{coef_WF_LMTO})] with the overlap
matrix $O_{nn'}({\bf k})$ is defined as
\begin{eqnarray}
\label{O-S1} O_{nn'}({\bf k}) & = & \langle \widetilde{W}_{n\bf
k}|\widetilde{W}_{n'\bf k}\rangle = \sum_{\mu } \tilde{b}_{\mu
n}^{{\bf k}*} \tilde{b}_{\mu n'}^{\bf k}. \nonumber
\end{eqnarray}
The occupancy matrix in the orthogonalized WF basis [Eq.~(\ref{WF1})] is defined as
\begin{eqnarray}
\label{Q_WF4} Q_{nn'}({\bf T})& = & -\frac{1}{\pi} {\rm Im}
\int_{-\infty}^{E_f} d\varepsilon \int \int d{\bf r} d{\bf r
}' \frac{1}{N}\sum_{\bf k} W^*_{n{\bf k}}({\bf r}) G^{\bf k}({\bf r},{\bf
r}',\varepsilon) W_{n'{\bf k}}({\bf r}')e^{-i\bf kT}.
\end{eqnarray}
By using Eq.~(\ref{Green1}) and orthogonalized Wannier functions [Eq.~(\ref{WF1})], one finds
\begin{eqnarray}
Q_{nn'}({\bf T})=
\frac{1}{N}\sum_{\bf k} \sum_{\mu \nu} b_{\mu n}^{{\bf k}*}  b_{\nu n'}^{\bf k}
  Q^{\bf k}_{\mu\nu}e^{-i\bf kT},
\end{eqnarray}
with
\begin{eqnarray}
 Q^{\bf k}_{\mu\nu} & = & -\frac{1}{\pi} {\rm Im} \int_{-\infty}^{E_f}
d\varepsilon G^{\bf k}_{\mu \nu}(\varepsilon).
\end{eqnarray}

The energy matrix can be defined similarly (except that the
integral over energy is calculated in the $(E_1,E_2)$ interval
where the corresponding WFs are determined) as
\begin{eqnarray}
\label{E_WF4} E_{nn'}(\bf T)& = & -\frac{1}{\pi} {\rm Im}
\int_{E_1}^{E_2}\varepsilon d\varepsilon \int \int d{\bf r}
d{\bf r}' \frac{1}{N}\sum_{\bf k} W^*_{n{\bf k}}({\bf r}) G^{\bf
k}({\bf
r},{\bf r}',\varepsilon) W_{n'{\bf k}}({\bf r}')e^{-i\bf kT}  \\
&=&
\frac{1}{N}
\sum_{\bf k} \sum_{\mu \nu} b_{\mu n}^{{\bf k}*}  b_{\nu n'}^{\bf k}
  E^{\bf k}_{\mu\nu}e^{-i\bf kT},  \nonumber
\end{eqnarray}
with
\begin{eqnarray}
 E^{\bf k}_{\mu\nu} & = & -\frac{1}{\pi} {\rm Im} \int_{E_1}^{E_2}
\varepsilon d\varepsilon G^{\bf k}_{\mu \nu}(\varepsilon).
\end{eqnarray}

\section{DMFT for ODF fluctuation Hamiltonian}
\label{WF_DMFT}

The DMFT\cite{vollha93,pruschke,georges96} was recently found to be a
powerful tool to numerically solve multiband Hubbard models. In order to use this tool 
the fluctuation Hamiltonian [Eq.~(\ref{H_corr3})] should be rewritten in the form of standard 
multi-orbital Hubbard model. For that one needs to identify Coulomb parameters $U_{nn'}$ 
as derivatives $\partial E_{n}/\partial Q_{n'}$ and rearrange terms 
in Eq.~(\ref{H_corr3}) into noninteracting and interaction parts:
\begin{eqnarray}
\label{H_corr4} \widehat{H}_{ODF} = \widehat{H}^{0}+\widehat{H}_{int}
\end{eqnarray}
with interaction term:
\begin{eqnarray}
\label{H_corr5} \widehat{H}_{int} & = & 
\frac{1}{2}\sum_{n}\sum_{n'\neq n}U_{nn'}\widehat{Q}_n\widehat{Q}_{n'}
\end{eqnarray}
and noninteracting Hamiltonian:
\begin{eqnarray}
\label{H_corr6} \widehat{H}^{0} & = & \widehat{H}^0_{ODF}
+\frac{1}{2}\sum_{n}\sum_{n'\neq n}U_{nn'}\langle\widehat{Q}_n\rangle\langle\widehat{Q}_{n'}\rangle -\sum_{n}\widehat{Q}_n\sum_{n'\neq n}U_{nn'}\langle\widehat{Q}_{n'}\rangle.
\end{eqnarray}

In DMFT the lattice problem becomes an effective single-site
problem which has to be solved self-consistently
for the matrix self energy ${\widehat{ \Sigma}} $ and the local
matrix Green function in Wannier functions basis set:
\begin{eqnarray}
\label{gloc} G_{nn'}(\varepsilon)=\!\frac{1}{V_{BZ}} \int d{\bf k}
\left( \left[ \;(\varepsilon-\mu)
\widehat{1}-\widehat{H}^{0}({\bf k})
-\widehat{\Sigma}(\varepsilon)\right]^{-1}\right)_{nn'},
\end{eqnarray}
where $\mu$ is a chemical potential, $\widehat{H}^{0}$ is noninteracting Hamiltonian 
[Eq.~(\ref{H_corr6})] and $\widehat{\Sigma}(\varepsilon)$ is self-energy in Wannier functions basis:
\begin{eqnarray}
\label{Sigma}
  \widehat{\Sigma}(\varepsilon)= \sum_{nn'} |W_{n} \rangle \Sigma_{nn'}(\varepsilon) \langle W_{n'}|.
\end{eqnarray}

The DMFT single-site problem may be viewed as a self-consistent
single-impurity Anderson model.~\cite{georges96} The corresponding
local one-particle matrix Green function ${\widehat{G}}$ can be
written as a functional integral\cite{georges96} involving an
action where the Hamiltonian of the correlation problem under
investigation, including the interaction term with the Hubbard
interaction, enters.~\cite{LDADMFT} The
action depends on the bath matrix Green function ${\widehat{\cal
{G}}}$ through
\begin{eqnarray}
\label{G0} {(\widehat{\cal {G}})}^{-1} = (\widehat{G})^{-1} +
{\widehat{\Sigma}}.
\end{eqnarray}
To solve the functional integral of the effective single-impurity Anderson
problem, various methods can be used: quantum Monte Carlo (QMC),
numerical renormalization group (NRG), exact diagonalization (ED),
non-crossing approximation (NCA), etc. (for a brief overview of
the methods see Ref.~\onlinecite{LDADMFT}). In the present work the QMC 
method was used to solve the impurity problem.

\section{Self-consistency}
\label{self-cons}

The matrix self-energy $\widehat\Sigma (\varepsilon)$ 
obtained as a solution of DMFT in Sec.~\ref{WF_DMFT} is defined in
the WF basis set [Eq.~(\ref{Sigma})]. In order to compute the
interacting Green function in the full-orbital Hilbert space
[Eqs.~(\ref{Green1}) and (\ref{Green2})] one has to transform self-energy to the
LMTO basis set. This can be easily done by using
the linear-expansion form of the WFs in terms of the LMTO 
basis set [Eqs.~(\ref{WF_orth}) and (\ref{coef_WF_LMTO})],
\begin{eqnarray}
\label{Sigma_full}
\Sigma^{\bf
k}_{\mu\nu}(\varepsilon)  & \equiv &  \langle \phi_{\mu}^{\bf k}|
\widehat{\Sigma}(\varepsilon)|\phi_{\nu}^{\bf k}\rangle
 =  \sum_{nn'}  \langle \phi_{\mu}^k|W_{n\bf k} \rangle
\Sigma_{nn'}(\varepsilon)    \langle W_{n'\bf k}|\phi_{\nu}^{\bf
k}\rangle = \sum_{nn'} b_{\mu n}^{\bf k} \Sigma_{nn'}(\varepsilon)
b_{\nu n'}^{{\bf k}*}.
\end{eqnarray}

The matrix elements of the self-energy $\Sigma^{\bf k}_{\mu\nu}
(\varepsilon)$ [Eq.~(\ref{Sigma_full})] together with the
noninteracting Hamiltonian matrix $H^0_{\mu\nu}(\bf k)$ 
[Eqs.~(\ref{H_corr6}) and (\ref{H_constrain2})] allow one
to calculate the matrix Green function $G^{\bf
k}_{\mu\nu}(\varepsilon)$ [Eq.~(\ref{Green2})] and thus the 
interacting Green function $G({\bf r},{\bf r'},\varepsilon)$
[Eq.~(\ref{Green1})]. $G({\bf r},{\bf r'},\varepsilon)$ contains the
full information about the system, and various electronic,
magnetic, and spectral properties can be obtained from it. 

One can also calculate the charge-density distribution modified by
correlation effects via
\begin{eqnarray}
\label{rho}
 \rho(\bf r)&=& -\frac{1}{\pi} {\rm Im}
\int_{-\infty}^{E_f}d\varepsilon G({\bf r},{\bf
r},\varepsilon).
\end{eqnarray}
With this $\rho(\bf r)$ one can recalculate the LDA potential 
which is a functional of electron density [Eq.~(\ref{LDA49})]. From the 
Green function [Eq.~(\ref{Green1})] one can recalculate new WFs
[Eqs.~(\ref{WF1}) and (\ref{coef_green_b})] which together with the new LDA
Hamiltonian allows one to obtain new parameters for the
noninteracting Hamiltonian [Eqs.~(\ref{H_constrain2}) and (\ref{H_corr6})]. 
With the new set of Wannier functions one
performs a series of constrain LDA calculation to determine derivatives 
$\partial E_{n}/\partial Q_{n'}$ and hence define a new Coulomb interaction 
parameters $U_{nn'}$ in the interaction Hamiltonian [Eq.~(\ref{H_corr5})]. 
The set of new LDA potential, WFs, and Coulomb interaction parameters
calculated from the interacting Green function [Eq.~(\ref{Green1})]
defines the input for the next iteration step and hence closes the
self-consistency loop in the computation scheme. For the
feedback from DMFT to LDA in the particular case of the LMTO
method\cite{LMTO} one needs a set of moments for the partial
densities of states $M^{(p)}_{ql}$ for every atomic sphere $q$ and
the orbital moment $l$ (Ref.~\onlinecite{Skriver84}) in order to calculate the
new charge density and hence the new LDA potential:
\begin{eqnarray}
\label{M-G}
 M^{(p)}_{ql} & = & \int_{-\infty}^{E_f}d\varepsilon
\varepsilon^p N_{ql}(\varepsilon), \\ \nonumber
N_{ql}(\varepsilon)&=& -\frac{1}{\pi N}{\rm Im} \sum_{\bf k} \sum_m
G^{\bf k}_{qlm,qlm}(\varepsilon).
\end{eqnarray}

The orbital densities functional computational scheme described above 
is {\it{ab initio}} (does not contain any outside parameters) and fully 
self-consistent. The charge density $\rho(\bf r)$, Wannier functions 
$W_{n}({\bf r})$, and derivatives $\partial E_{n}/\partial Q_{n'}$ 
(effective fluctuations interaction strength) are recalculated on every 
self-consistency loop and hence modified by correlations comparing with 
the values obtained in standard LDA calculations. Analogous calculation scheme 
was proposed in Fig.~1 in Ref.~\onlinecite{WF-DMFT}.

In the present work ODF method was applied to the problem of metal-insulator 
transition in nonstoichiometric lanthanum trihydride LaH$_{3-x}$, 
where both parts of the calculation scheme, static mean-field approximation 
[Eqs.~(\ref{functional-GTS})-(\ref{dV-GTS})] 
and fluctuation Hamiltonian [Eq.~(\ref{H_corr3})] in DMFT (Sec. \ref{WF_DMFT}) 
were needed to describe experimentally observed dependence 
of the system ground state on the hydrogen deficiency parameter $x$.

\section{Electronic structure of nonstoichiometric lanthanum trihydride }
\label{LaH3}
\begin{figure}
  \includegraphics[clip=true,width=0.45\textwidth]{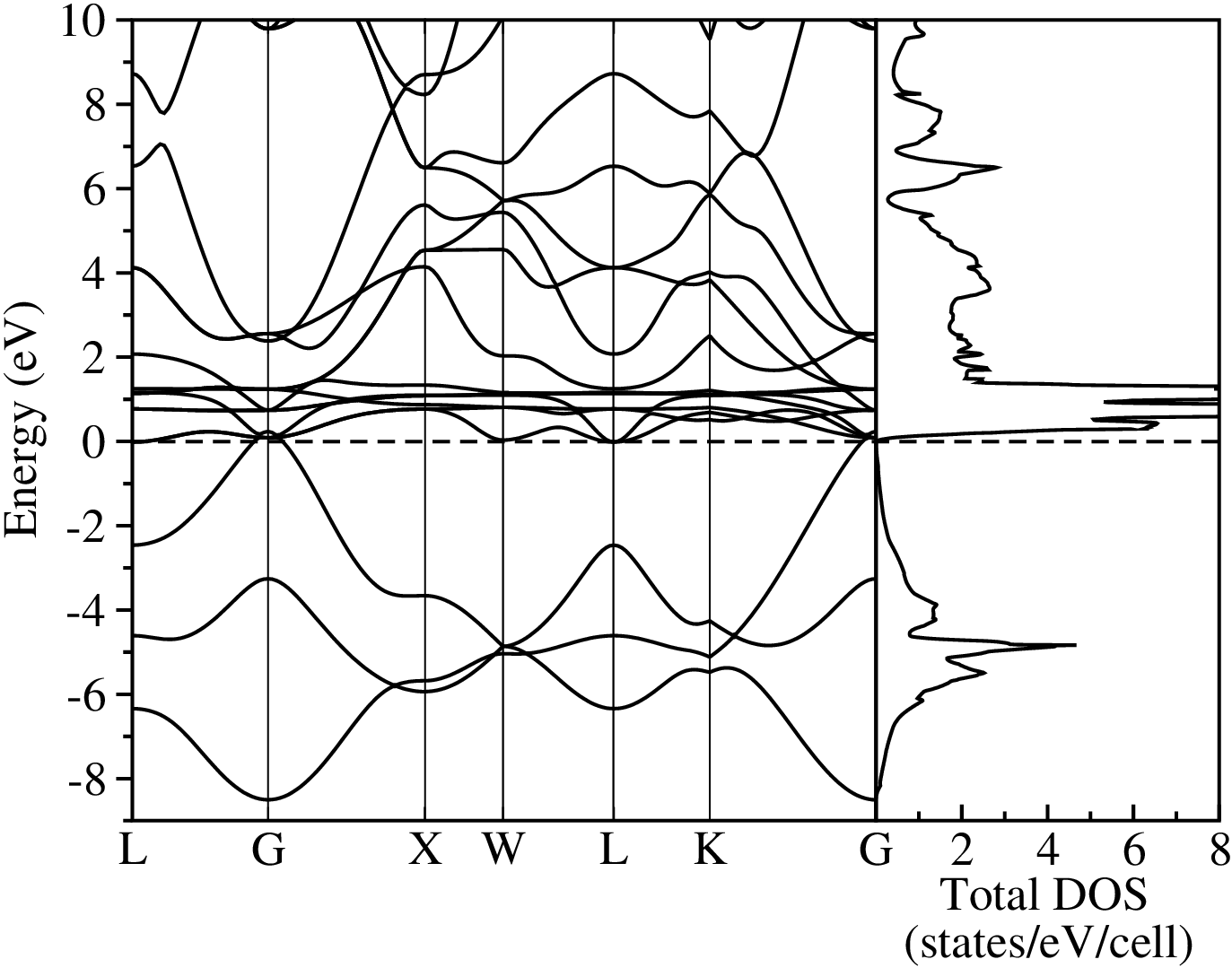}
  \caption{LaH$_3$ band structure and density of states calculated in standard LDA.}
  \label{LaH3_lda}
\end{figure}

Lanthanum trihydride LaH$_{3-x}$ shows an interesting metal-insulator transition 
with increasing of $x$ value.~\cite{LaH3-prop} While stoichiometric LaH$_3$ is an insulator, 
30\% hydrogen deficiency ($x\approx0.3$) results in a metallic ground state. 
However, standard LDA calculations\cite{LaH3-LDA} give a metal as a ground state 
even for stoichiometric composition LaH$_{3}$. 
In this case well-known problem of LDA, underestimation of the energy gap value, 
is so severe that the gap value is negative with valence and conduction bands overlapping 
on 0.25 eV (see Fig.~\ref{LaH3_lda}). 

The lanthanum trihydride has a crystal structure derived from the face-centered cubic structure 
(lattice parameter $a$ = 10.5946 a.u.) for La atoms and hydrogen atoms occupying 
two tetrahedral and one octahedral interstitials per every metallic ion. The band structure 
of LaH$_3$ (Fig.~\ref{LaH3_lda}) is relatively simple: lower three occupied bands are formed 
by hydrogen $1s$ states and the conduction bands correspond to lanthanum 5$d$, 6$s$, and 6$p$ states.

Attempts to cure the LDA fault in LaH$_3$ were performed with many methods beyond LDA, 
among others, the GW method\cite{LaH3-GW} (the energy gap $E_g$=0.8-0.9 eV), the weighted local density 
approach\cite{LaH3-WDA} ($E_g$=0.7 eV), and model calculations.~\cite{LaH3-Model} 

\begin{figure}
  \includegraphics[clip=true,width=0.45\textwidth]{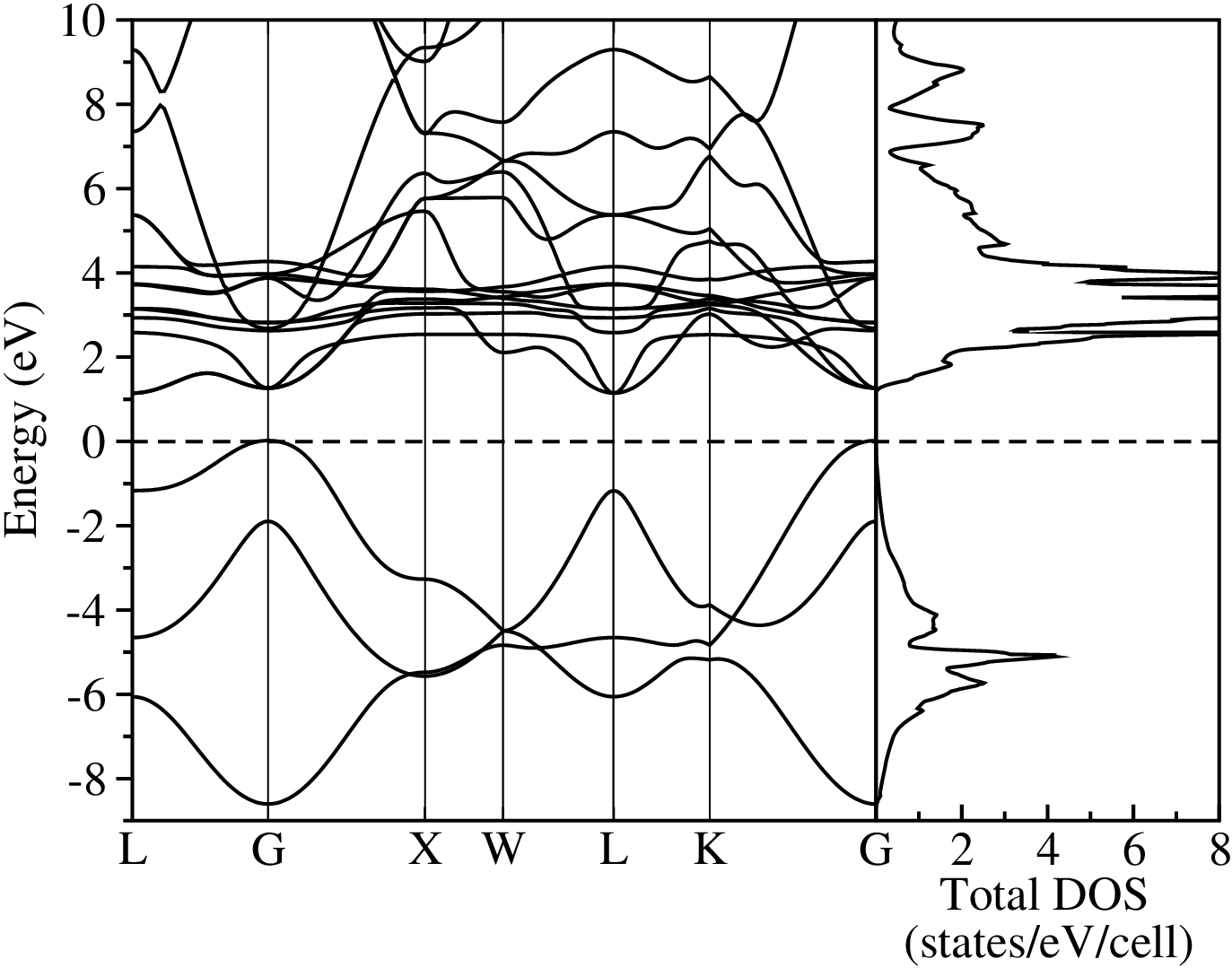}
  \caption{LaH$_3$ band structure and density of states calculated in orbital densities functional (ODF) theory [Eq.~(\ref{functional-GTS})]. Zero of energy is at the Fermi energy.}
  \label{LaH3_gts}
\end{figure}

The orbital densities functional (ODF) theory [Eq.~(\ref{functional-GTS})] proposed 
in the present work adds to standard Kohn-Sham equations potential 
correction in the form of projection operator [Eqs.~(\ref{H_corr}) and (\ref{dV-GTS})]. 
This correction is negative for occupied valence bands and positive for empty conduction 
bands and hence increases energy separation between those bands. ODF calculations 
for LaH$_3$ results in insulating ground state with a fundamental gap value of 1.10 eV and 
direct optical gap 1.25 eV at G-point (see Fig.~\ref{LaH3_gts}). Experimental data for the value 
of energy gap E$_g$=0.5 eV was estimated using activation energy determined from resistivity 
measurements in Ref.~\onlinecite{gap}. Optical measurements estimate the direct gap 
E$_g$ as 1.87 eV but the fundamental band gap $\sim$1 eV lower.~\cite{LaH3-gap}

\begin{figure}
  \includegraphics[clip=true,width=0.45\textwidth]{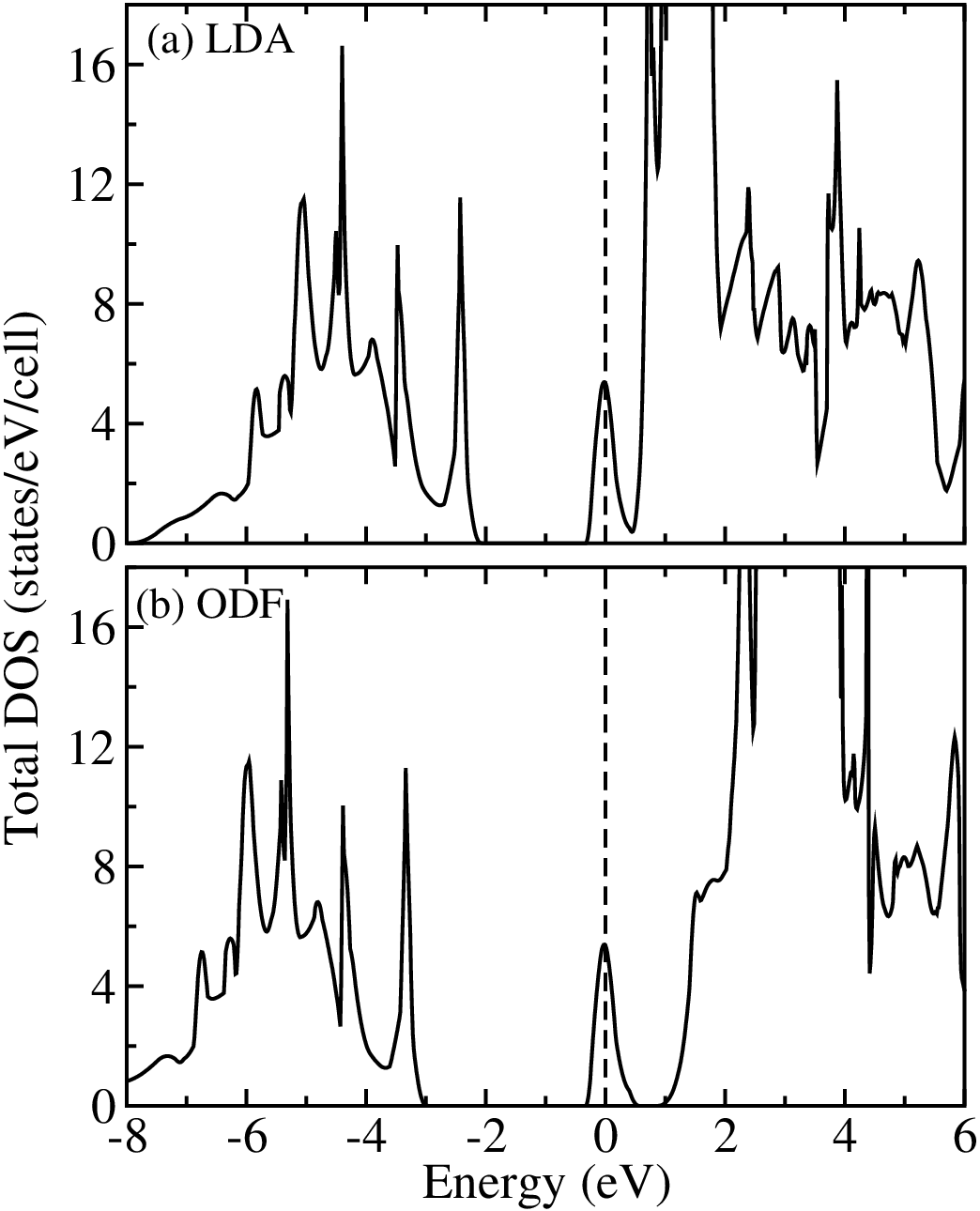}
  \caption{LaH%
$_{2.75}$ density of states calculated (a) in standard LDA and (b) in orbital densities functional (ODF) theory [Eq.~(\ref{functional-GTS})].}
  \label{La4H11v}
\end{figure}

\begin{figure}
  \includegraphics[clip=true,width=0.45\textwidth]{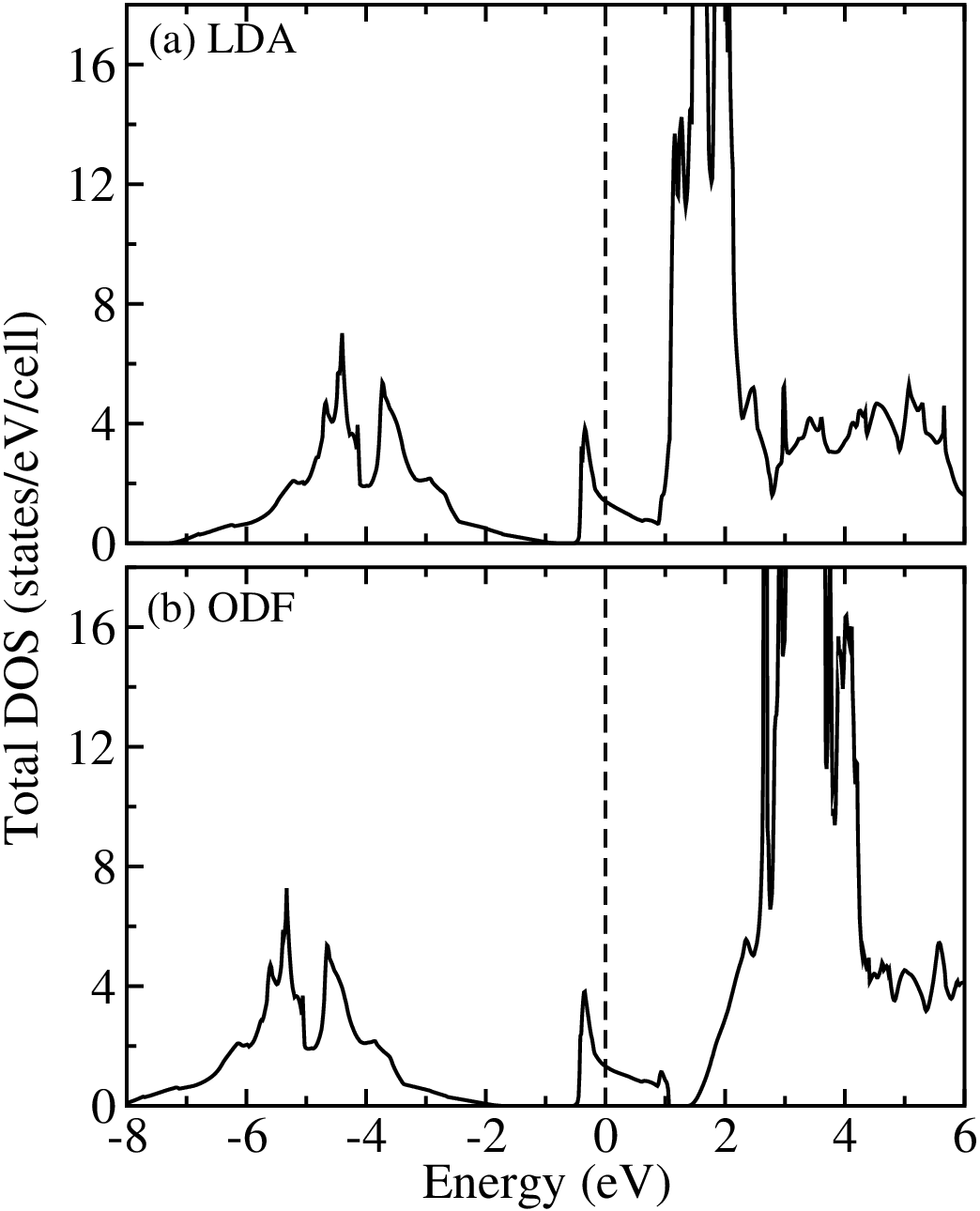}
  \caption{LaH%
$_{2.5}$ density of states calculated (a) in standard LDA and (b) in orbital densities functional (ODF) theory [Eq.~(\ref{functional-GTS})]. }
  \label{La2H5v}
\end{figure}

Hydrogen atoms could be removed from the lanthanum trihydride forming nonstoichiometric LaH$_{3-x}$. 
We have performed LDA and ODF calculations for two compositions $x$=0.25 (LaH$_{2.75}$) 
(see Fig.~\ref{La4H11v}) and $x$=0.5 (LaH$_{2.5}$) (see Fig.~\ref{La2H5v}).

\begin{figure}
  \includegraphics[clip=true,width=0.45\textwidth]{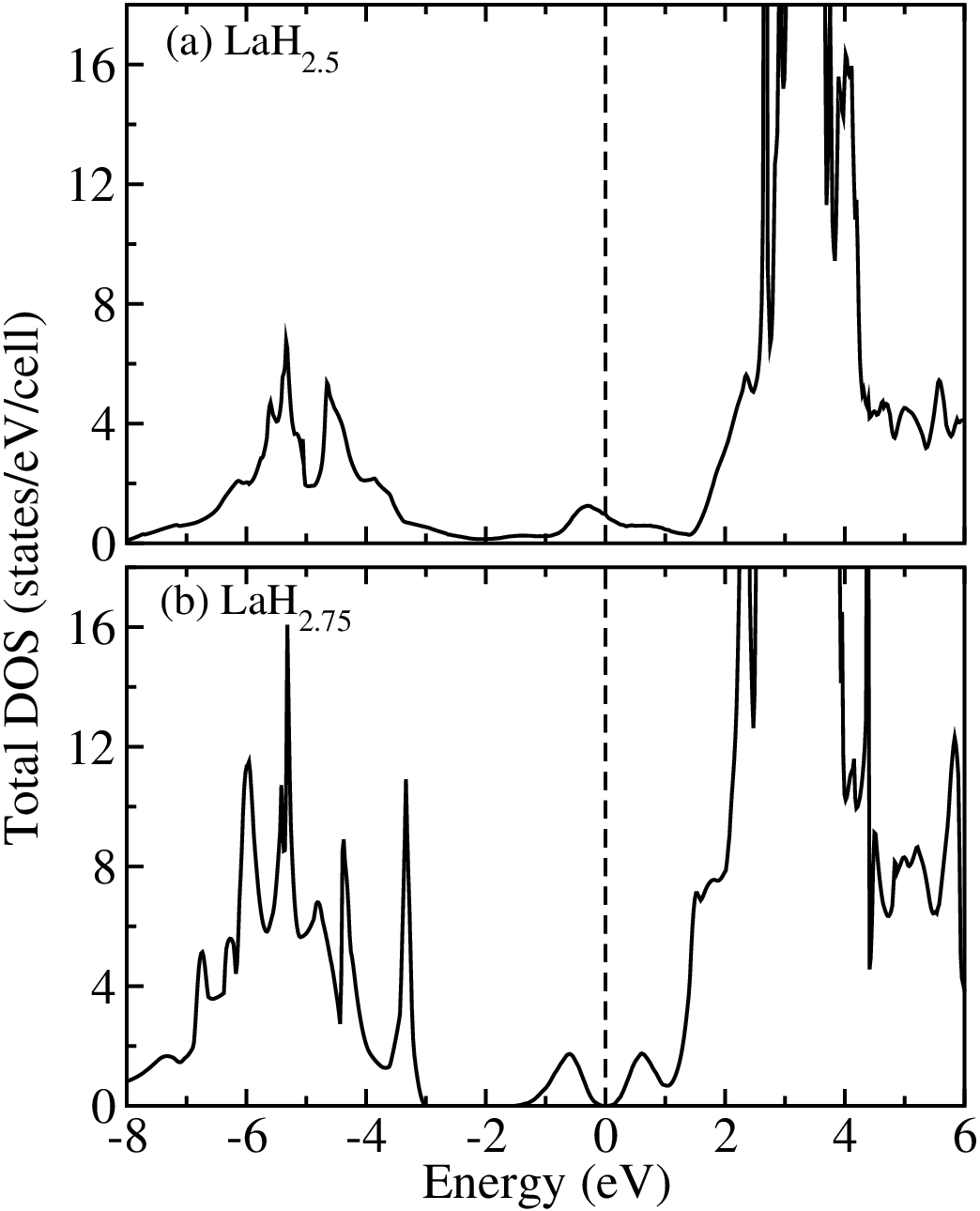}
  \caption{(a) LaH%
$_{2.5}$ and (b) LaH$_{2.75}$ densities of states calculated with ODF fluctuation 
  Hamiltonian [Eq.~(\ref{H_corr3})] in DMFT-QMC.}
  \label{dmft}
\end{figure}

Both LDA and ODF calculations gave metallic ground state for two compositions $x$=0.25 and $x$=0.5. 
Every hydrogen vacancy leads to appearance of one vacancy state in the gap split from 
the conduction band with one electron occupying this state (see Fig.~\ref{La4H11v}(a)). 
Static mean-field ODF potential correction [Eqs.~(\ref{H_corr}) and (\ref{dV-GTS})] results 
in separation of the vacancy band from the conduction band and in increasing of energy gap 
between valence and conduction bands comparing with LDA but half-filled vacancy band stays metallic 
(see Figs.~\ref{La4H11v}(b) and~\ref{La2H5v}(b)). However, experimentally 
only LaH$_{2.5}$ is metallic while LaH$_{2.75}$ should be an insulator. To treat this problem 
we have used ODF fluctuation Hamiltonian [Eq.~(\ref{H_corr3})] and have solved it using 
DMFT-QMC (see Sec. \ref{WF_DMFT}). Inverse temperature 
parameter was $\beta$=10~eV$^{-1}$. In the result we have obtained 
in agreement with experiment paramagnetic insulator ground state for LaH$_{2.75}$ 
with a typical Mott insulator pattern of lower and upper Hubbard bands around chemical potential with 
the energy gap of 0.1~eV (see Fig.~\ref{dmft}(b)). However, for LaH$_{2.5}$ a width 
of the vacancy band is significantly larger then for LaH$_{2.75}$ (see Figs.~\ref{La4H11v}(b) 
and \ref{La2H5v}(b)) and DMFT calculations resulted in well defined metallic state 
(see Fig.~\ref{dmft}(a)).

\section{Conclusion}
\label{Conclusion}
We have proposed orbital densities functional that by construction 
has discontinuous exchange-correlation potential in agreement 
with \emph{exact} density functional theory. The one-electron potential 
obtained by variation of this functional produces comparing with LDA lower 
energies for valence bands and higher energies for conduction bands thus 
overcoming systematic underestimation of energy gap value in LDA. 
To treat correlation effects we have derived from orbital densities functional 
the Hamiltonian corresponding to the fluctuations of orbital densities around
the ground state values. Combining this Hamiltonian with dynamical mean-filed theory 
and quantum Monte Carlo method for effective impurity problem we have 
developed \emph{ab initio} and fully self-consistent scheme for electronic 
structure calculations. The scheme was applied to the problem of metal-insulator 
transition in LaH$_{3-x}$ and gave a qualitative improvement for agreement 
with experimental data.

\section{Acknowledgments}
This work was supported by Russian Foundation for Basic Research
under the grants RFFI-04-02-16096 and RFFI-03-02-39024, by Netherlands Organization for Scientific
Research through NWO 047.016.005. A.L. acknowledges support from the Dynasty 
Foundation and International Centre for Fundamental Physics in Moscow. 
We are grateful to D.~Vollhardt 
for helpful discussions and for supplying QMC-DMFT computer code.

\appendix
\section{Constrain LDA functional and its derivatives}
\label{constrain}
LDA functional $E_{LDA}[\rho]$ depends on the electron density
$\rho(\bf{r})$ defined by one-electron functions $\psi_i(\mathbf{r})$
and their occupancies $n_i$:
\begin{eqnarray}
\label{LDA43} \rho(\mathbf{r}) = \sum_i n_i|\psi_i(\bf{r})|^2,
\end{eqnarray}
\begin{eqnarray}
\label{LDA44} E_{LDA}[\rho]  = T[\rho]   +
\frac{1}{2}\int d\mathbf{r}\int
d\mathbf{r'} \frac{\rho(\mathbf{r})\rho(\mathbf{r'})}{|\mathbf{r}-\mathbf{r'}|}   \\
\nonumber + \int d\mathbf{r} V_Z(\mathbf{r})\rho(\mathbf{r}) +
E_{xc}[\rho],
\end{eqnarray}
where $V_Z(\mathbf{r})$ is external potential, $E_{xc}[\rho]$ is exchange-correlation
energy functional [see Eq.~(\ref{ODF2})] and $T[\rho]$ is kinetic energy term.

Kinetic energy is a functional of electron density $T[\rho]$ via single particle 
orbitals $\psi_i(\mathbf{r})$ which are functionals of density themselves. 
To prove that the following reasoning can be used.~\cite{gross} 
Hohenberg-Kohn theorem\cite{HK} states that external potential is uniquely defined 
by the ground state electron density of the system. That is also true for 
a noninteracting system with potential $V_S[\rho](\mathbf{r})$. This potential 
defines a Schr\"odinger equation for the orbitals $\psi_i(\mathbf{r})$:
\begin{eqnarray}
\label{LDA46} (-\nabla^2+V_S[\rho](\mathbf{r}))\psi_i(\mathbf{r}) =
\epsilon_i\psi_i(\mathbf{r}).
\end{eqnarray}
As potential $V_S[\rho](\mathbf{r})$ is a functional of density then solutions 
of the Eq.~(\ref{LDA46}) are also functionals of density:
\begin{eqnarray}
\label{LDA47} \psi_i(\mathbf{r}) =\psi_i[\rho](\mathbf{r}).
\end{eqnarray}
 
Kinetic energy is defined by a set of orbitals $\psi_i[\rho](\mathbf{r})$ and hence it is also a functional of density:
\begin{eqnarray}
\label{LDA45} T[\rho]  =  \sum_i n_i\int d\mathbf{r}
\psi^{*}_i[\rho](\mathbf{r})(-\nabla^2)\psi_i[\rho](\mathbf{r}).  
\end{eqnarray}

Minimization of the functional Eq.~(\ref{LDA44}) gives:
\begin{eqnarray}
\label{LDA48} 
\frac{\delta
E_{LDA}[\rho]}{\delta\rho(\mathbf{r})} = \frac{\delta
T[\rho]}{\delta\rho(\mathbf{r})} +V_{LDA}(\mathbf{r}) = 0
\end{eqnarray}
where one-electron LDA potential is:
\begin{eqnarray}
\label{LDA49} V_{LDA}(\mathbf{r}) = \int d\mathbf{r'}
\frac{\rho(\mathbf{r'})}{|\mathbf{r}-\mathbf{r'}|} +
V_Z(\mathbf{r}) + V_{xc}(\mathbf{r})
\end{eqnarray}
($V_{xc}(\mathbf{r})=\delta E_{xc}[\rho]/\delta\rho(\mathbf{r})$
is exchange-correlation potential).

Variation of kinetic energy functional is:
\begin{eqnarray}
\label{LDA50} \delta T[\rho] & = & \delta \Big( \sum_i n_i\int d\mathbf{r}
\psi^{*}_i[\rho](\mathbf{r})(-\nabla^2)\psi_i[\rho](\mathbf{r})\Big) \\ \nonumber
& = & \delta \Big( \sum_i n_i\int d\mathbf{r}
\psi^{*}_i[\rho](\mathbf{r})(\epsilon_i[\rho]-V_S[\rho](\mathbf{r}))\psi_i[\rho](\mathbf{r})\Big)\\ \nonumber & = & 
\delta \Big(\sum_i n_i\epsilon_i[\rho]-\int d\mathbf{r} V_S[\rho](\mathbf{r})\rho(\mathbf{r})\Big) .
\end{eqnarray}
First order perturbation theory for one-electron energies $\epsilon_i[\rho]$ gives:
\begin{eqnarray}
\label{LDA51} \delta \epsilon_i[\rho] & = & \int d\mathbf{r}
\psi^{*}_i[\rho](\mathbf{r})(\delta V_S[\rho](\mathbf{r}) )\psi_i[\rho](\mathbf{r})\Big .
\end{eqnarray}
Then Eq.~(\ref{LDA50}) is:
\begin{eqnarray}
\label{LDA52} \delta T[\rho]  =  
- \int d\mathbf{r'} V_S[\rho](\mathbf{r})\delta \rho(\mathbf{r}) 
\end{eqnarray}
and variational derivative of kinetic energy functional is:
\begin{eqnarray}
\label{LDA53s} 
 \frac{\delta
T[\rho]}{\delta\rho(\mathbf{r})} = -V_S[\rho](\mathbf{r}).
\end{eqnarray}
The minimization condition [Eq.~(\ref{LDA48})] becomes:
\begin{eqnarray}
\label{LDA53} 
 \frac{\delta
E_{LDA}[\rho]}{\delta\rho(\mathbf{r})} =  -V_S[\rho](\mathbf{r})+V_{LDA}(\mathbf{r}) = 0 .
\end{eqnarray}

A set of Kohn-Sham\cite{Kohn-Sham} equations (\ref{LDA46}) with $V_S[\rho](\mathbf{r}) = V_{LDA}(\mathbf{r})$ defines one-electron orbitals $\psi_i[\rho](\mathbf{r})$ and hence a new electron density $\rho(\mathbf{r})$ via Eq.~(\ref{LDA43}). The problem should be solved self-consistently till electron density used to calculate $V_{LDA}(\mathbf{r})$ will coincide with the density obtained using Kohn-Sham equations solutions $\psi_i[\rho](\mathbf{r})$.

In order to define functional [Eq.~(\ref{LDA44})] for charge density $\widetilde{\rho}(\mathbf{r})$ different from the density minimizing $E_{LDA}[\rho]$ one needs to introduce constrain potential $\Phi(\mathbf{r})$ and minimize auxiliary functional:
\begin{eqnarray}
\label{LDA41} F[\rho]  = E_{LDA}[\rho]  + \int d\mathbf{r} \Phi(\mathbf{r})(\rho(\mathbf{r})-\widetilde{\rho}(\mathbf{r})),
\end{eqnarray}
\begin{eqnarray}
\label{LDA42} \frac{\delta F}{\delta\rho(\mathbf{r})} =
\frac{\delta
E_{LDA}}{\delta\rho(\mathbf{r})}+\Phi(\mathbf{r})= -V_S[\rho](\mathbf{r})+V_{LDA}(\mathbf{r})+\Phi(\mathbf{r})=0 .
\end{eqnarray}
Then potential in Kohn-Sham equations (\ref{LDA46}) is:
\begin{eqnarray}
\label{LDA54} 
V_S[\rho](\mathbf{r})=V_{LDA}(\mathbf{r})+ \Phi(\mathbf{r}).
\end{eqnarray}
The second variational derivative of LDA functional is:
\begin{eqnarray}
\label{LDA55} 
 \frac{\delta^2
E_{LDA}[\rho]}{\delta\rho(\mathbf{r})\delta\rho(\mathbf{r'})} = \frac{\delta}{\delta\rho(\mathbf{r'})}(-V_S[\rho](\mathbf{r})+V_{LDA}(\mathbf{r}))=
\frac{\delta V_{LDA}(\mathbf{r})}{\delta\rho(\mathbf{r'})}-\frac{\delta V_S[\rho](\mathbf{r})}{\delta\rho(\mathbf{r'})}.
\end{eqnarray}
The variational derivative of LDA potential [Eq.~(\ref{LDA49})] is obtained in a straightforward way:
\begin{eqnarray}
\label{LDA56} 
\frac{\delta V_{LDA}(\mathbf{r})}{\delta\rho(\mathbf{r'})}= \frac{1}{|\mathbf{r}-\mathbf{r'}|} + \delta(\mathbf{r}-\mathbf{r'})\frac{\delta V_{xc}(\mathbf{r})}{\delta\rho(\mathbf{r'})}.
\end{eqnarray}
The second term in Eq.~(\ref{LDA55}), $\delta V_S[\rho](\mathbf{r})/\delta\rho(\mathbf{r'})$ is more complicated. 
Let us first solve an inverse problem -- to calculate response function:~\cite{levy}
\begin{eqnarray}
\label{LDA57s} 
\chi(\mathbf{r},\mathbf{r'}) \equiv \frac{\delta\rho(\mathbf{r})}{\delta V_S(\mathbf{r'})}.
\end{eqnarray}
First-order perturbation theory for eigenfunctions and eigenvalues of Kohn-Sham equations (\ref{LDA46}) gives:
\begin{eqnarray}
\label{LDA58s} 
\frac{\delta\psi_i(\mathbf{r})}{\delta V_S(\mathbf{r'})}=\sum_{j\neq i}\psi_j(\mathbf{r})\frac{\psi_j^*(\mathbf{r'})\psi_i(\mathbf{r'})}{\epsilon_i-\epsilon_j},\\
\frac{\delta\epsilon_i}{\delta V_S(\mathbf{r'})}=\psi_i^*(\mathbf{r'})\psi_i(\mathbf{r'}).
\end{eqnarray}
Then response function in Eq.~(\ref{LDA57s}) is:
\begin{eqnarray}
\label{LDA57} 
\chi(\mathbf{r},\mathbf{r'}) \equiv \frac{\delta\rho(\mathbf{r})}{\delta V_S(\mathbf{r'})} = \sum_i n_i
\sum_{j\neq i}\psi_j(\mathbf{r})\frac{\psi_i^*(\mathbf{r})\psi_j(\mathbf{r})\psi_j^*(\mathbf{r'})\psi_i(\mathbf{r'})}{\epsilon_i-\epsilon_j} + c.c.
\end{eqnarray}
The second term in Eq.~(\ref{LDA55}) is defined by the inverse response function:
\begin{eqnarray}
\label{LDA58} 
\frac{\delta V_S(\mathbf{r})}{\delta\rho(\mathbf{r'})}= \chi^{-1}(\mathbf{r},\mathbf{r'}).
\end{eqnarray}
This inverse response function can be obtained solving integral equation:
\begin{eqnarray}
\label{LDA59} 
 \int d\mathbf{r''}\chi(\mathbf{r},\mathbf{r''}) \chi^{-1}(\mathbf{r''},\mathbf{r'})=\delta(\mathbf{r}-\mathbf{r'}).
\end{eqnarray}
Then the second variational derivative of LDA functional is:
\begin{eqnarray}
\label{LDA60} 
 \frac{\delta^2
E_{LDA}}{\delta\rho(\mathbf{r})\delta\rho(\mathbf{r'})} = \frac{1}{|\mathbf{r}-\mathbf{r'}|} + \delta(\mathbf{r}-\mathbf{r'})\frac{\delta V_{xc}(\mathbf{r})}{\delta\rho(\mathbf{r'})} - \chi^{-1}(\mathbf{r},\mathbf{r'})\equiv W(\mathbf{r},\mathbf{r'}),
\end{eqnarray}
$W(\mathbf{r},\mathbf{r'})$ can be interpreted as an effective interaction strength between density fluctuations $\delta\rho(\mathbf{r}),\delta\rho(\mathbf{r'})$ in LDA functional.
The first variational derivative of the LDA functional can be obtained from Eq.~(\ref{LDA42}):
\begin{eqnarray}
\label{LDA61} 
\frac{\delta
E_{LDA}}{\delta\rho(\mathbf{r})}= -\Phi[\rho](\mathbf{r}),
\end{eqnarray}
where $\Phi[\rho](\mathbf{r})$ is constrain potential needed to obtain 
electron density $\rho(\mathbf{r})$ in self-consistent solution 
of Kohn-Sham equations [Eqs.~(\ref{LDA46}) and (\ref{LDA54})].

Let us consider variation of the LDA functional Eq.~(\ref{LDA44}) in respect 
to the occupancies $n_i$ and orbitals $\psi_i(\mathbf{r})$ defining electron density 
$\rho(\mathbf{r})$ in Eq.~(\ref{LDA43}):
\begin{eqnarray}
\label{LDA62} 
\delta
E_{LDA}= \sum_i \Big(\delta n_i \frac{\delta
E_{LDA}}{\delta n_i} + \int d\mathbf{r}\delta\psi_i(\mathbf{r})\frac{\delta
E_{LDA}}{\delta\psi_i(\mathbf{r})} \Big) ,
\end{eqnarray}

\begin{eqnarray}
\label{LDA63} 
  \frac{\delta
E_{LDA}}{\delta n_i}& = & \frac{\delta T}{\delta n_i} + \int d\mathbf{r}\frac{\delta (E_{LDA}-T)}{\delta\rho(\mathbf{r})}\frac{\delta\rho(\mathbf{r})}{\delta n_i}  \\ \nonumber
& = & \int d\mathbf{r}\psi^{*}_i(\mathbf{r})(-\nabla^2)\psi_i(\mathbf{r}) + \int d\mathbf{r}\psi^{*}_i(\mathbf{r})V_{LDA}(\mathbf{r})\psi_i(\mathbf{r})
 \\ \nonumber
& = & \int d\mathbf{r}\psi^{*}_i(\mathbf{r})(-\nabla^2+V_{LDA}(\mathbf{r}))\psi_i(\mathbf{r}) = \epsilon_i,
\end{eqnarray}

\begin{eqnarray}
\label{LDA64} 
  \frac{\delta
E_{LDA}}{\delta \psi_i(\mathbf{r})}& = & n_i(-\nabla^2)\psi^*_i(\mathbf{r}) +  n_iV_{LDA}(\mathbf{r})\psi^*_i(\mathbf{r})  \\ \nonumber
& = & n_i(-\nabla^2+V_{LDA}(\mathbf{r}))\psi^*_i(\mathbf{r}) = n_i  \epsilon_i\psi^*_i(\mathbf{r}),
\end{eqnarray}
\begin{eqnarray}
\label{LDA65} 
\delta
E_{LDA}= \sum_i \Big(\delta n_i\epsilon_i + n_i\epsilon_i\int d\mathbf{r}\delta\psi_i(\mathbf{r})\psi^*_i(\mathbf{r}) \Big) .
\end{eqnarray}
 As orbitals are normalized:
 \begin{eqnarray}
\label{LDA66} 
\int d\mathbf{r}\delta\psi_i(\mathbf{r})\psi^*_i(\mathbf{r}) = \delta (\int d\mathbf{r}\psi_i(\mathbf{r})\psi^*_i(\mathbf{r})-1)=0
\end{eqnarray}
and from Eq.~(\ref{LDA65}) it follows that:
\begin{eqnarray}
\label{LDA67} 
\frac{\delta E_{LDA}}{\delta n_i}=\epsilon_i .
\end{eqnarray}

All this was proven to be valid for the LDA eigenfunctions
$\psi_i(\mathbf{r})$ which are solutions of Kohn-Sham equations
[Eq.~(\ref{LDA46})]. Let us prove that the derivative of LDA
functional over the Wannier function occupancy $Q_{n}$ [Eq.~(\ref{Q_WF})] 
is equal to the energy $E_{n}$ [Eq.~(\ref{E_WF})] of this WF.

The charge density expressed via Wannier functions $W_n({\bf r})$
[Eq.~(\ref{WF_orth})] and its occupancies $Q_{n}$ is:
\begin{eqnarray}
\label{LDA11} \rho(\mathbf{r}) = \sum_{\mathbf{T}}\sum_n
Q_{n}|W_n({\mathbf r}-\mathbf{T})|^2.
\end{eqnarray}
Then derivative over Wannier function occupancy $Q_{n}$ (with fixed functions 
$W_n({\mathbf r})$) can be calculated using Eq.~(\ref{LDA63}):
\begin{eqnarray}
\label{LDA13} \frac{\delta E_{LDA}}{\delta Q_{n}}  =   \int d\mathbf{r}
W^*_n({\mathbf r})(-\nabla^2+V(\mathbf{r}))W_n({\mathbf r}) = \int
d\mathbf{r} W^*_n({\mathbf r})\widehat{H}_{LDA}W_n({\mathbf
r})=E_{n}.
\end{eqnarray}

\section{Projection operator in ODF functional}
\label{project}
 In ODF functional Eq.~(\ref{functional-GTS})
Wannier functions occupancies $Q_{n}$ [Eq.~(\ref{Q_WF})] can
be expressed via projection operator $\widehat{Q}_n$:
\begin{eqnarray} \label{Q_WF2}
Q_{n} & = & \sum_{i}
    \langle\psi_{i}|\widehat{Q}_n|\psi_{i}\rangle,
    \\ \nonumber
 \widehat{Q}_n &  = & |W_{n}^{\bf T}\rangle\langle W_{n}^{\bf T}|
\end{eqnarray}
(summation $\sum_{i}$ is over occupied states only in all
equations below).
 The variational derivative of occupancy
$Q_{n}$ with respect to the one-electron wave function $\delta\psi_i(\mathbf{r})$ is:
\begin{eqnarray}
\label{LDA18} \frac{\delta Q_{n}}{\delta\psi_i(\mathbf{r})} =
\widehat{Q}_n \psi^*_{i}(\mathbf{r}).
\end{eqnarray}
Then variation of the correction term in ODF functional Eq.~(\ref{functional-GTS}) 
will give the following correction to the
corresponding expression for LDA functional [see Eq.~(\ref{LDA64})]:
\begin{eqnarray}
\label{LDA19}\frac{\delta E_{ODF}}{\delta\psi_i(\mathbf{r})} & = &
\frac{\delta E_{LDA}}{\delta\psi_i(\mathbf{r})} -\sum_n
\frac{\partial E_{n}}{\partial
Q_{n}}(Q_{n}-\frac{1}{2})\frac{\delta
Q_{n}}{\delta\psi_i(\mathbf{r})}\\ \nonumber & = &
(-\nabla^2+V_{LDA}(\mathbf{r}))\psi^*_i(\mathbf{r}) + \sum_n
\frac{\partial E_{n}}{\partial
Q_{n}}(\frac{1}{2}-Q_{n})\widehat{Q}_n\psi^*_{i}(\mathbf{r})\\
\nonumber & = & (-\nabla^2+V_{LDA}(\mathbf{r})+\sum_n \delta
V_{n}\widehat{Q}_n)\psi^*_i(\mathbf{r}).
\end{eqnarray}
This proves Eqs.~(\ref{H_corr}) and (\ref{dV-GTS}) for ODF correction
to LDA Hamiltonian.

\section{Total energy in LDA and ODF functionals }
Kinetic energy term in Eq.~(\ref{LDA44}) can be calculated using
Eq.~(\ref{LDA46}):
\begin{eqnarray}
\label{LDA14} (-\nabla^2)\psi_i(\mathbf{r}) =( \epsilon_i
-V_{LDA}(\mathbf{r}))\psi_i(\mathbf{r}).
\end{eqnarray}
Multiplying Eq.~(\ref{LDA14}) by $\psi^*_i(\mathbf{r})$,
integrating over $\mathbf{r}$ and summing over occupied states $i$
one can obtain:
\begin{eqnarray}
\label{LDA15} \sum_i\int
d\mathbf{r}\psi^*_i(\mathbf{r})(-\nabla^2)\psi_i(\mathbf{r})
=\sum_i\epsilon_i - \int d\mathbf{r} V_{LDA}(\mathbf{r})\rho(\mathbf{r}).
\end{eqnarray}
Then total energy in Eq.~(\ref{LDA44}) is equal to:
\begin{eqnarray}
\label{LDA16} E_{LDA}  = \sum_i\epsilon_i - \int d\mathbf{r}
V_{LDA}(\mathbf{r})\rho(\mathbf{r})   + \frac{1}{2}\int d\mathbf{r}\int
d\mathbf{r'} \frac{\rho(\mathbf{r})\rho(\mathbf{r'})}{|\mathbf{r}-\mathbf{r'}|}  \\
\nonumber + \int d\mathbf{r} V_Z(\mathbf{r})\rho(\mathbf{r}) +
\int d\mathbf{r}
\varepsilon_{xc}(\rho(\mathbf{r}))\rho(\mathbf{r}).
\end{eqnarray}
Using Eq.~(\ref{LDA49}) for LDA potential $V(\mathbf{r})$ the
final expression for total energy is:
\begin{eqnarray}
\label{LDA17} E_{LDA}  = \sum_i\epsilon_i   - \frac{1}{2}\int
d\mathbf{r}\int d\mathbf{r'}
\frac{\rho(\mathbf{r})\rho(\mathbf{r'})}{|\mathbf{r}-\mathbf{r'}|}
+  \int d\mathbf{r}
(\varepsilon_{xc}(\rho(\mathbf{r}))-V_{xc}(\rho(\mathbf{r})))\rho(\mathbf{r}).
\end{eqnarray}

In the case of ODF functional [Eq.~(\ref{functional-GTS})] there
will be an additional term in expression for kinetic energy [Eq.~(\ref{LDA15})] 
due to the ODF potential correction [Eq.~(\ref{LDA19})]:
\begin{eqnarray}
\label{LDA20} -\sum_i\int d\mathbf{r}\psi^*_i(\mathbf{r})(\sum_n
\delta V_{n}\widehat{Q}_n)\psi_i(\mathbf{r}) = -\sum_n \delta
V_{n}Q_{n}.
\end{eqnarray}
The additional terms to total energy expression Eq.~(\ref{LDA17})
are
\begin{eqnarray}
\label{LDA21} E_{ODF}& = & \tilde{E}_{LDA}- \sum_n \delta
V_{n}Q_{n}-\frac{1}{2} \sum_n \frac{\partial
E_{n}}{\partial Q_{n}}Q_{n}(Q_{n}-1) \\
\nonumber & = & \tilde{E}_{LDA} + \frac{1}{2} \sum_n
\frac{\partial E_{n}}{\partial Q_{n}}(Q_{n})^2.
\end{eqnarray}
Here $\tilde{E}_{LDA}$ is defined as right part of Eq.~(\ref{LDA17}). 
$\tilde{E}_{LDA}$ differs from $E_{LDA}$ because it is taken 
for $\rho({\mathbf{r}})$ which does not minimize $E_{LDA}$.



\newpage

\end{document}